\documentclass[12pt]{elsart}

\usepackage{graphicx}
\usepackage{cmap}
\usepackage[plainpages=false,pdfpagelabels,pdftex,bookmarks=true,bookmarksnumbered=false,colorlinks=true,linkcolor=blue,citecolor=blue,urlcolor=blue]{hyperref}
\hypersetup{
  pdfauthor = {James Rohlf},
  pdftitle = {Radiation Damage Studies of Silicon Photomultipliers},
  pdfsubject = {silicon photomultiplier},
  pdfkeywords = {QIE, HCAL, CMS, SiPM, quantum efficiency},
  bookmarksopen=true,
  bookmarksopenlevel=1
}

\usepackage{verbatim}

\begin{document}
\begin{frontmatter} 

\title{Radiation Damage Studies of Silicon Photomultipliers}

\author{P. Bohn, A. Clough, E. Hazen, A. Heering, and J. Rohlf} 
\address{Department of Physics, Boston University, Boston, MA 02212, USA} 
\author{J. Freeman and S. Los}
\address{Fermi National Accelerator Laboratory, Batavia, IL 60510, USA}
\author{E. Cascio}
\address{Francis H. Burr Proton Therapy Center, Massachusetts General Hospital, Boston, MA 02114}
\author{S. Kuleshov}  \footnote{Also at Departamento de Fisica y Centro de Estudios Subatomicos,
Universidad Tecnica Federico Santa Maria, Casilla 110-V, Valparaiso, Chile}    

\address{Institute for Theoretical and Experimental Physics, Moscow, Russia}
\author{Y. Musienko} \footnote{On leave from INR, Moscow, Russia}
\address{Department of Physics, Northeastern University, Boston, MA 02115, USA}
\author{C. Piemonte}
\address{Fondazione Bruno Kessler, Trento, I-38050, Italy}
\bigskip

\clearpage

\begin{abstract}
 We report on the measurement of the radiation hardness of silicon photomultipliers (SiPMs) manufactured by 
 Fondazione Bruno Kessler in Italy (1 mm$^2$ and 6.2 mm$^2$), Center of Perspective
Technology and Apparatus in Russia (1 mm$^2$ and 4.4 mm$^2$), and Hamamatsu Corporation in Japan (1 mm$^2$).
The SiPMs were irradiated using a beam of 212 MeV protons at Massachusetts General Hospital,
receiving fluences of up to $3 \times 10^{10}$ protons per cm$^2$ with the SiPMs at operating voltage.
Leakage currents were read continuously during the irradiation.
The delivery of the protons was paused periodically to record scope traces in response to calibrated light pulses to monitor the gains, photon detection efficiencies, and dark counts of the SiPMs.
The leakage current and dark noise are found to increase with fluence.
Te leakage current is found to be proportional to the mean square deviation of the noise distribution, indicating the dark counts are due to increased random individual pixel activation, while SiPMs remain fully functional as photon detectors. 
The SiPMs are found to anneal at room temperature with a reduction in the leakage current by a factor of 2 in about 100 days.

\end{abstract}
\end{frontmatter}


Please send proofs to:\\
James W. Rohlf\\
Physics Department\\
Boston University\\
Boston, MA, USA\\
tel.: +1617-353-2600, fax: +1617-353-9393, email: rohlf@bu.edu
\bigskip
\bigskip

\section{Introduction}
During the last several years, we have investigated the use of silicon photomultipliers (SiPMs)~\cite{sipm1}-\cite{sipm2} 
to collect light from bundles of 1 mm fibers optically connected to the scintillators of the hadron calorimeter of the Compact Muon Solenoid (CMS)~\cite{cms} at the Large Hadron Collider (LHC)~\cite{lhc}. 
The SiPMs developed for use at the LHC must be sufficiently radiation hard to withstand the expected fluence.
Damage in silicon detectors depends on the flux, type and energy of the particles. The damage produced by protons depends on their energy-dependant non-ionizing energy losses (NIEL). 
For LHC detectors, particle fluxes have been calculated in 1 MeV neutron equivalent fluxes. The damage produced by the 212 MeV protons used for these measurements is about 0.8 of that produced by 1 MeV neutrons~\cite{lind}.
The fluence for one LHC lifetime in the proximity of the CMS hadron outer (HO) photodetectors is expected to be approximately equivalent to $10^{10}$ per cm$^2$~\cite{mika}-\cite{julie}. While many silicon devices have been proven to be robust under LHC fluences~\cite{rd50}, no previous measurements are available for the latest generation of SiPMs with an active area ($A$) of several mm$^2$.

The SiPMs chosen for irradiation were  $A=6.2$ mm$^2$ round diodes from Fondazione Bruno Kessler (FBK, formerly ICT-irst) in Italy,  and $A =$ 4.4 mm$^2$ square diodes from
the Center of Perspective
Technology and Apparatus (CPTA) in Russia.  We also irradiated $A= $ 1.0 mm$^2$ square diodes from FBK, CPTA, and 
Hamamatsu Corporation (HC) in Japan. All of the SiPMs have a pixel size of 50 $\mu$m $\times$ 50 $\mu$m~\cite{sipm2}. In addition, we made measurements of a single pixel on an FBK 6.2 mm$^2$ SiPM. 

\section{Experimental Setup}

\begin{figure}[htbp]
  \centering
  \includegraphics[width=1\textwidth]{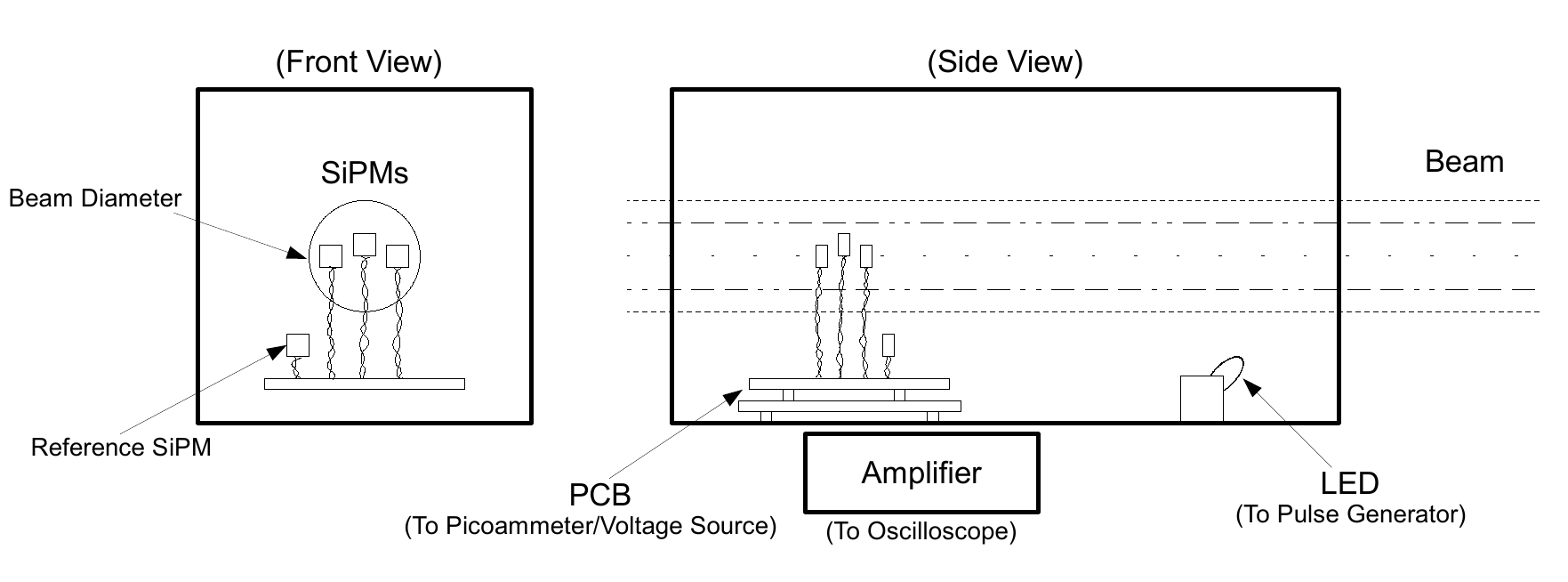}
  \caption{Layout of the SiPMs in the dark box. Three SiPMs were exposed to protons simultaneously, while a fourth reference SiPM was placed out of the beam. The currents of the SiPMs were read out continuously during exposure. The LED was pulsed before and after the exposure.}
\label{fig:box}
\end{figure}

\begin{figure}[htbp]
  \centering
  \includegraphics[width=0.7\textwidth]{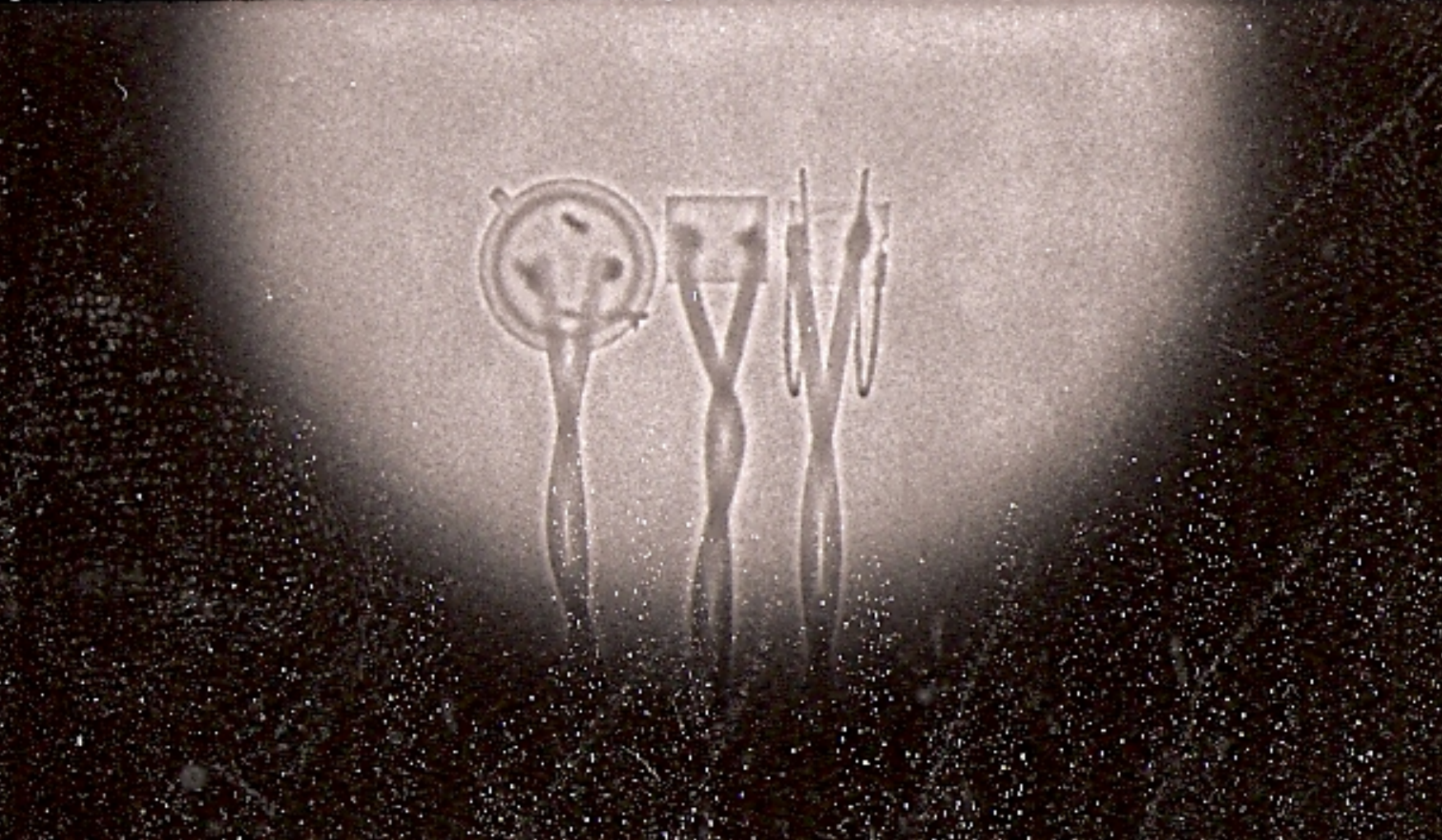}
  \caption{Exposure of the proton beam to a sheet of polaroid film was used to check positioning of the SiPMs.}
\label{fig:pol}
\end{figure}

The radiation studies were carried out at the  proton cyclotron~\cite{mgh} at the Massachusetts General Hospital Francis H. Burr Proton Therapy Center in Boston, MA, USA. The proton kinetic energy at the SiPMs was 212 MeV. The beam spot size was 4 cm diameter, allowing irradiation of three SiPMs simultaneously.
The fluence delivered on target was measured directly during irradiation using a thin-foil transmission ion chamber whose response was calibrated to the fluence with a Faraday cup.  The SiPMs were mounted in groups of 4 on printed circuit boards. 
The SiPM boards were mounted in a dark box together with a light-emitting diode (LED) as indicated in fig.~\ref{fig:box}.
The three SiPMs to be irradiated were extended vertically above the circuit boards into the proton beam by their electrical leads.  A fourth    
CPTA 4.4 mm$^2$ SiPM was mounted on each circuit board out of the radiation area and monitored 
before, during, and after irradiation as a reference diode. 
The positioning of the SiPM within the beam profile was checked directly with a photographic film exposure as shown in fig.~\ref{fig:pol}

\begin{figure}[htbp]
  \centering
  \includegraphics[width=1\textwidth]{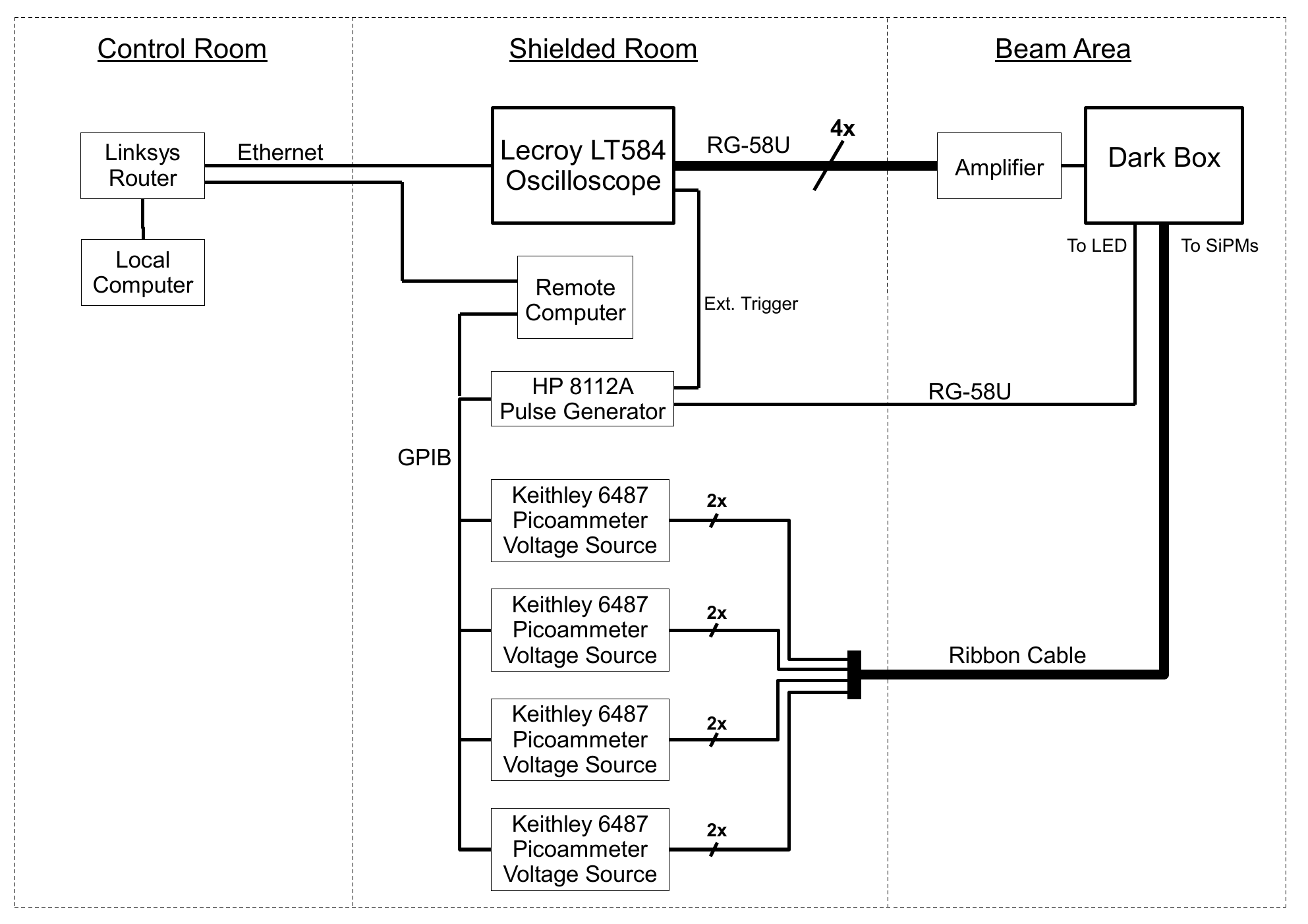}
  \caption{Layout of the SiPM readout. The digital oscilloscope was viewed and set remotely from the control room out of the radiation zone. The oscilloscope and power supplies were located in a shielded room several meters from the beam area where the SiPMs were mounted in a dark box.}
\label{fig:setup}
\end{figure}

Figure~\ref{fig:setup} shows a block diagram of the readout.
The nominal operating voltage ($V_{\rm b}$) was set individually for each SiPM to be approximately 3 V above turn-on (zero current) for the CPTA and FBK devices
using Keithley 6487 power supplies.
 The gain in this region of $V_{\rm b}$ was measured to be linear, varying from about 20 fC/PE per V for CPTA 4.4 mm$^2$ to 200 fC/PE per V for FBK 6.2 mm$^2$.  The resulting leakage current ($I_{\rm b}$) per SiPM active area was in the range of 1-2 $\mu$A/mm$^2$. 
The HC SiPMs have a steeper $I_{\rm b}$ {\it vs.} $V_{\rm b}$ curve and were set to be about 1 V above turn-on, resulting in a leakage current per area of about 0.1 $\mu$A/mm$^2$.

The SiPMs were mounted on 4 different circuit boards and were irradiated as summarized in Table 1. Boards 1 and 2 were populated with like SiPMs  
(CPTA 4.4 mm$^2$ reference, CPTA 1.0 mm$^2$, HC 1.0 mm$^2$, and FBK 1.0 mm$^2$)
and were exposed to fluences of $10^{10}$ protons per cm$^2$ for board 1 and $3\times 10^{10}$ protons per cm$^2$ for board 2.
Similarly boards 3 and 4 were populated with the same types of SiPMs
(CPTA 4.4 mm$^2$ reference, CPTA 4.4 mm$^2$, FBK 6.2 mm$^2$, and FBK single pixel) and irradiated to $10^{10}$ protons per cm$^2$ for board 3 and $3\times 10^{10}$ protons per cm$^2$ for board 4.
Boards 1 and 3 were irradiated in steps of $2.5 \times 10^9$ protons per cm$^2$ up to a total fluence of $10^{10}$ protons per cm$^2$. The fluence for each step was delivered uniformly over a time of 5 minutes. Several minutes were taken between irradiation steps in order to record waveforms.
Boards 2 and 4 were irradiated in steps of $2.5 \times 10^9$ protons per cm$^2$ up to a partial fluence of $10^{10}$ protons per cm$^2$ and then further exposed with two more steps of 
$10^{10}$ protons per cm$^2$, for a total fluence of $3\times 10^{10}$ protons per cm$^2$.
The SiPMs were kept at nominal operating voltage during the irradiation to allow continuous monitoring of $I_{\rm b}$.

\bigskip
\begin{table}[h]
\begin{center}
\caption{Irradiated SiPMs, nominal operating voltages, and proton fluences. }

\bigskip
\begin{tabular}[]{|c|c|c|c|}
\hline
{\it Board } & {\it SiPM}  & {$V_{\rm b}$ (V)}  & {\it Fluence \rm(cm$^{-2}$)}  \\
\hline
{1} & {\rm CPTA 4.4 mm$^2$ reference} & {36} & {$0$} \\
\hline
{1} & {\rm CPTA 1.0 mm$^2$}  & {34} &{$10^{10}$} \\
\hline
{1} & {\rm HC 1.0 mm$^2$}  & {70.5} &  {$10^{10}$} \\
\hline
{1} & {\rm FBK 1.0 mm$^2$}  & {33.5}   &{$10^{10}$} \\
\hline
\hline
{2} & {\rm CPTA 4.4 mm$^2$ reference} & {35} & {$ 0$} \\
\hline
{2} & {\rm CPTA 1.0 mm$^2$}  &{34} &{$ 3\times 10^{10}$} \\
\hline
{2} & {\rm HC 1.0 mm$^2$}  & {70.5} & {$ 3\times 10^{10}$} \\
\hline
{2} & {\rm FBK 1.0 mm$^2$} & {33.5}  & {$ 3\times 10^{10}$} \\
\hline
\hline
{3} & {\rm CPTA 4.4 mm$^2$ reference} & {35}  & {$0$} \\
\hline
{3} & {\rm CPTA 4.4mm$^2$} & {37}  & {$10^{10}$} \\
\hline
{3} & {\rm FBK 6.2 mm$^2$}& {34}  & {$10^{10}$} \\
\hline
{3} & {\rm FBK single pixel}& {37}   & {$10^{10}$} \\
\hline
\hline
{4} & {\rm CPTA 4.4 mm$^2$ reference}& {35}    & {$ 0$} \\
\hline
{4} & {\rm CPTA 4.4 mm$^2$}& {37}   & {$ 3\times 10^{10}$} \\
\hline
{4} & {\rm FBK 6.2 mm$^2$} & {34}  & {$ 3\times 10^{10}$} \\
\hline
{4} & {\rm FBK single pixel} & {37}  & {$ 3\times 10^{10}$} \\
\hline
\end{tabular}
\end{center}
\label{tab:det}
\end{table}
\bigskip

The LED was pulsed with a 50 MHz Hewlett-Packard 8112A pulse generator and files of 5000
waveforms were recorded with a Lecroy LT594 digital scope to monitor the pulse shape, signal, and noise distributions. The signal and noise distributions allowed monitoring of the gain ($M$) and number of photoelectrons ($n_{\rm PE}$).
The mean signal ($S$) in response to the LED may be written as
$$S = M n_{\rm PE} ~,$$
where the pedestal contribution due to electronic noise has been subtracted.
The root mean square (RMS) deviation ($\sigma$) from the mean may be written as
$$\sigma = M \sqrt{n_{\rm PE}F}~,$$
where the electronic noise has been subtracted in quadrature and
 $F$ is defined to be the excess noise factor, a number which has been independently measured to be close to unity for the SiPMs~\cite{sipm1}.
The measured distribution of $S$ then allows determination of the product of gain times excess noise factor,
$$MF = \frac{\sigma^2}{S} ~,$$
and the number of photoelectrons divided by the excess noise factor,
$$\frac{n_{\rm PE} }{F} = \frac{S^2}{\sigma^2} ~.$$

\section{Leakage Currents During Irradiation}

\begin{figure}[htbp]
  \centering
  \includegraphics[width=1\textwidth]{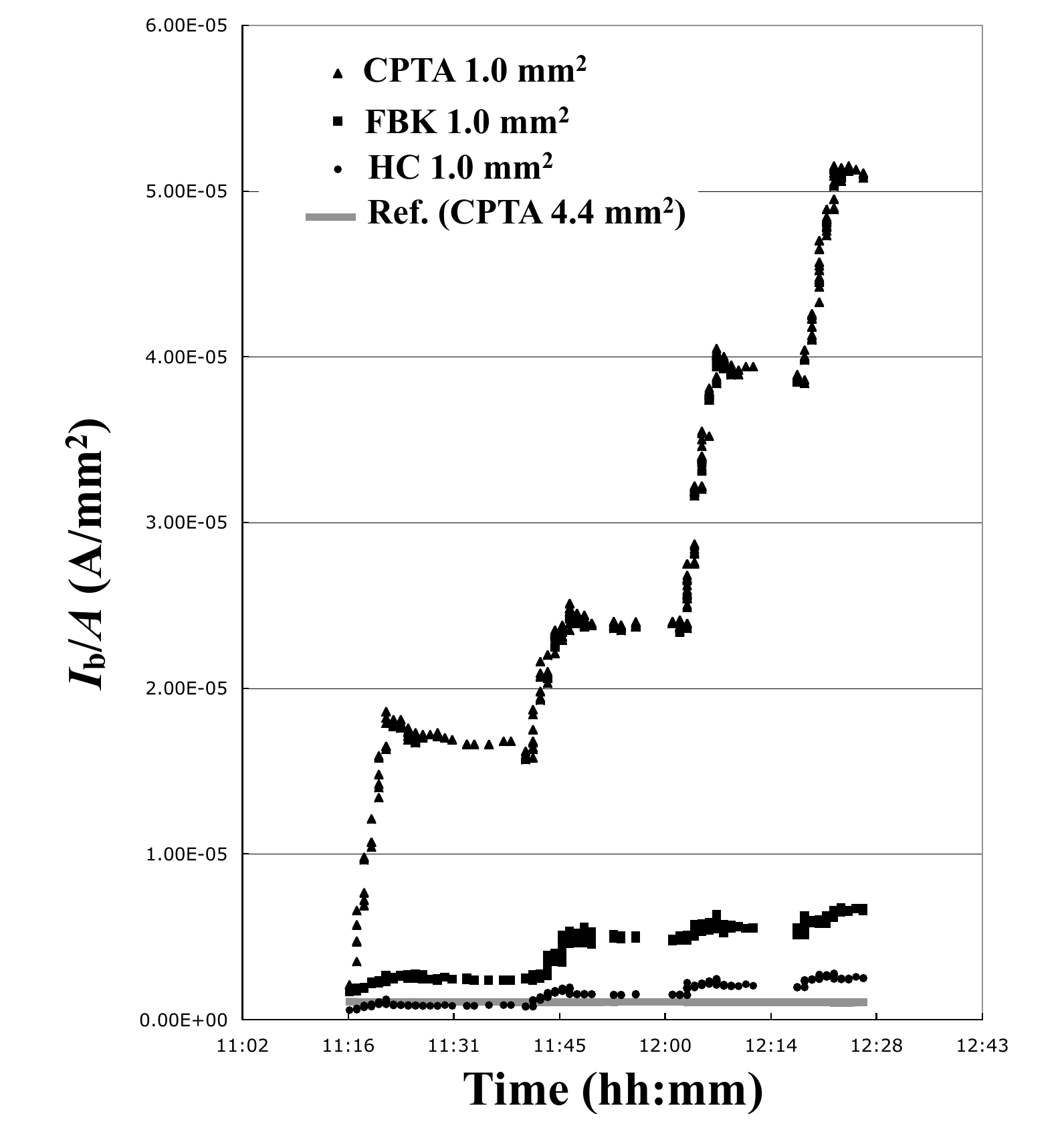}
  \caption{Leakage currents per area measured during irradiation for SiPMs on board 1:  CPTA 4.4 mm$^2$ reference diode (line), HC 1.0 mm$^2$ (circles), FBK 1.0 mm$^2$  (squares), and CPTA 1.0 mm$^2$ (triangles). The plateaus correspond to partial fluences of $2.5\times 10^9$, $5\times 10^9$, $7.5\times 10^9$, and $10^{10}$ protons per cm$^2$. A drop in leakage current due to room-temperature annealing is visible after each step.}
\label{fig:leak0}
\end{figure}

\begin{figure}[htbp]
  \centering
  \includegraphics[width=1\textwidth]{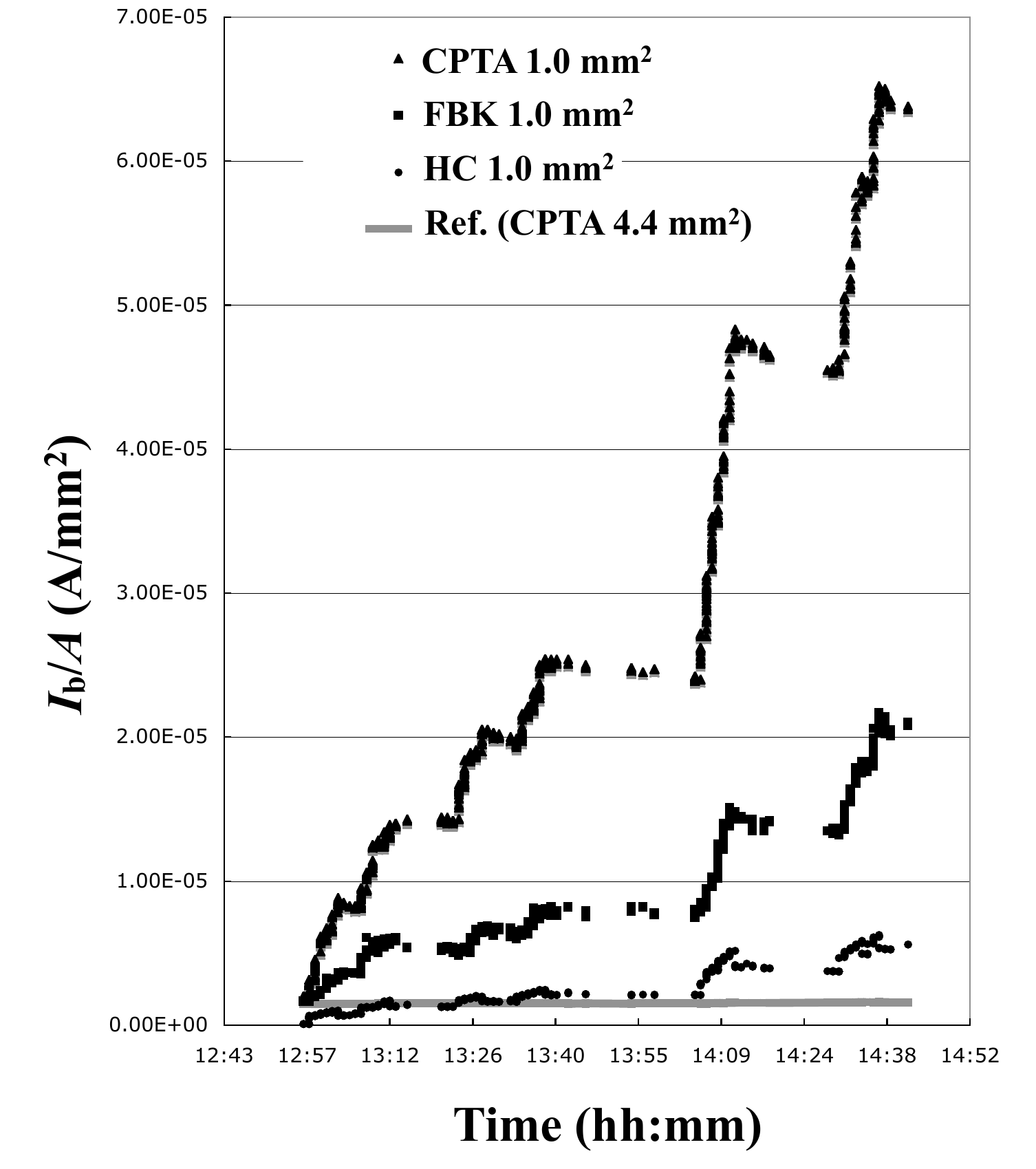}
  \caption{Leakage currents per area measured during irradiation for SiPMs on board 2:  CPTA 4.4 mm$^2$ reference diode (line), HC 1.0 mm$^2$ (circles), FBK 1.0 mm$^2$  (squares), and CPTA 1.0 mm$^2$ (triangles). The plateaus correspond to partial fluences of $2.5\times 10^9$, $5\times 10^9$, $7.5\times 10^9$, $10^{10}$, $2\times 10^{10}$, and $3\times 10^{10}$ protons per cm$^2$. A drop in leakage current due to room-temperature annealing is visible after each step.}
\label{fig:leak1}
\end{figure}

Leakage currents were read continuously during the irradiation. Figure~\ref{fig:leak0} shows $I_{\rm b}/A$   {\it vs.} time for the SiPMs on board 1. The plateaus correspond to partial fluences of $2.5\times 10^9$, $5\times 10^9$, $7.5\times 10^9$, and $10^{10}$ protons per cm$^2$, when the delivery of protons was paused in order to record SiPM waveforms. Figure~\ref{fig:leak1} shows $I_{\rm b}/A$  {\it vs.} time for the SiPMs on board 2. The plateaus correspond to partial fluences of $2.5\times 10^9$, $5\times 10^9$, $7.5\times 10^9$, $10^{10}$, $2\times 10^{10}$, and $3\times 10^{10}$ protons per cm$^2$. A drop in leakage current due to room-temperature annealing is visible after each irradiation step.

\begin{figure}[htbp]
  \centering
  \includegraphics[width=1\textwidth]{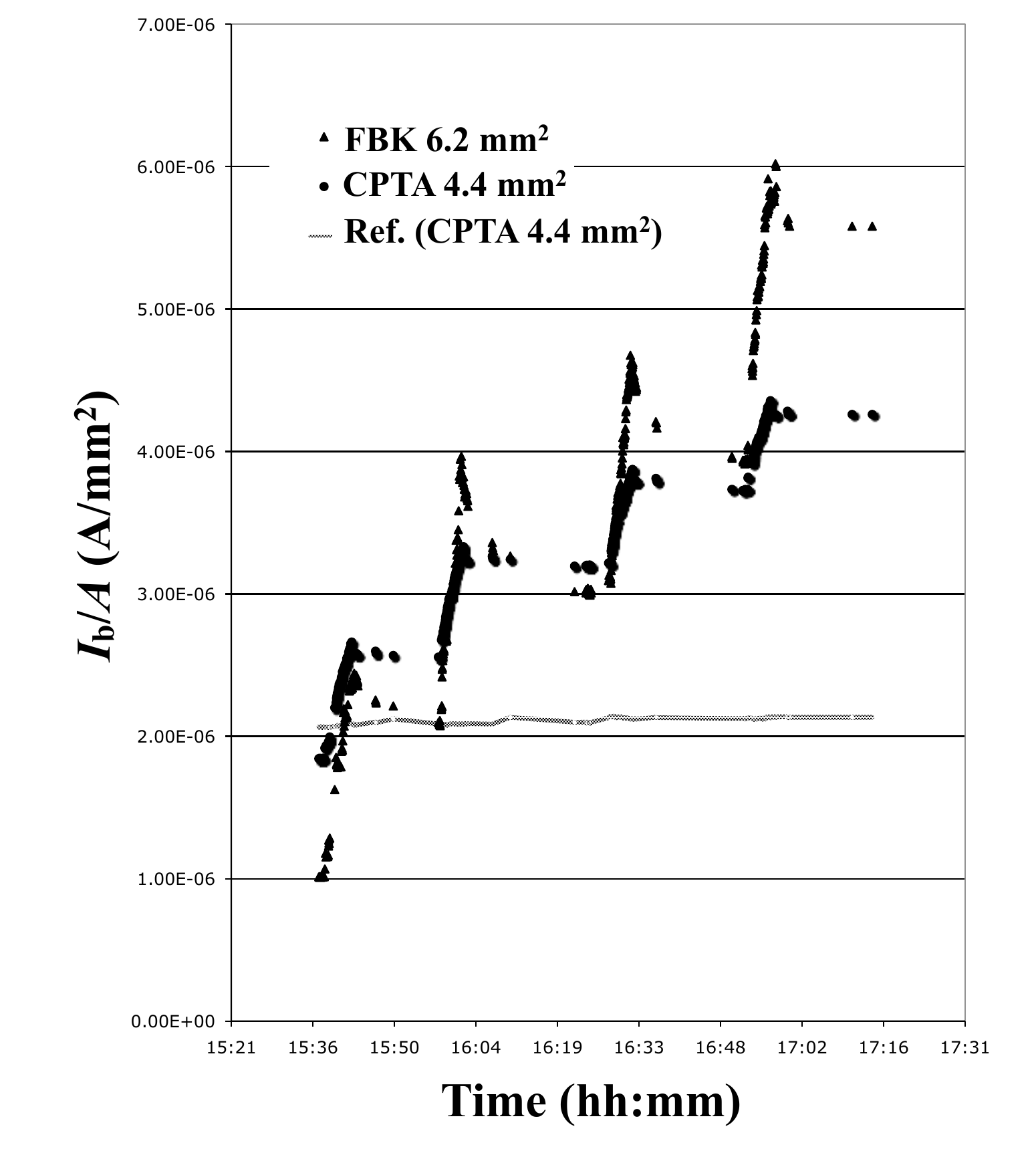}
  \caption{Leakage currents per area measured during irradiation for SiPMs on board 3:  CPTA 4.4 mm$^2$ reference diode (line), CPTA 4.4 mm$^2$ (circles), and FBK 6.2 mm$^2$  (triangles). The plateaus correspond to partial fluences of $2.5\times 10^9$, $5\times 10^9$, $7.5\times 10^9$, and $10^{10}$ protons per cm$^2$. A drop in leakage current due to room-temperature annealing is visible after each step.}
\label{fig:leak2} 
\end{figure}

\begin{figure}[htbp]
  \centering
  \includegraphics[width=1\textwidth]{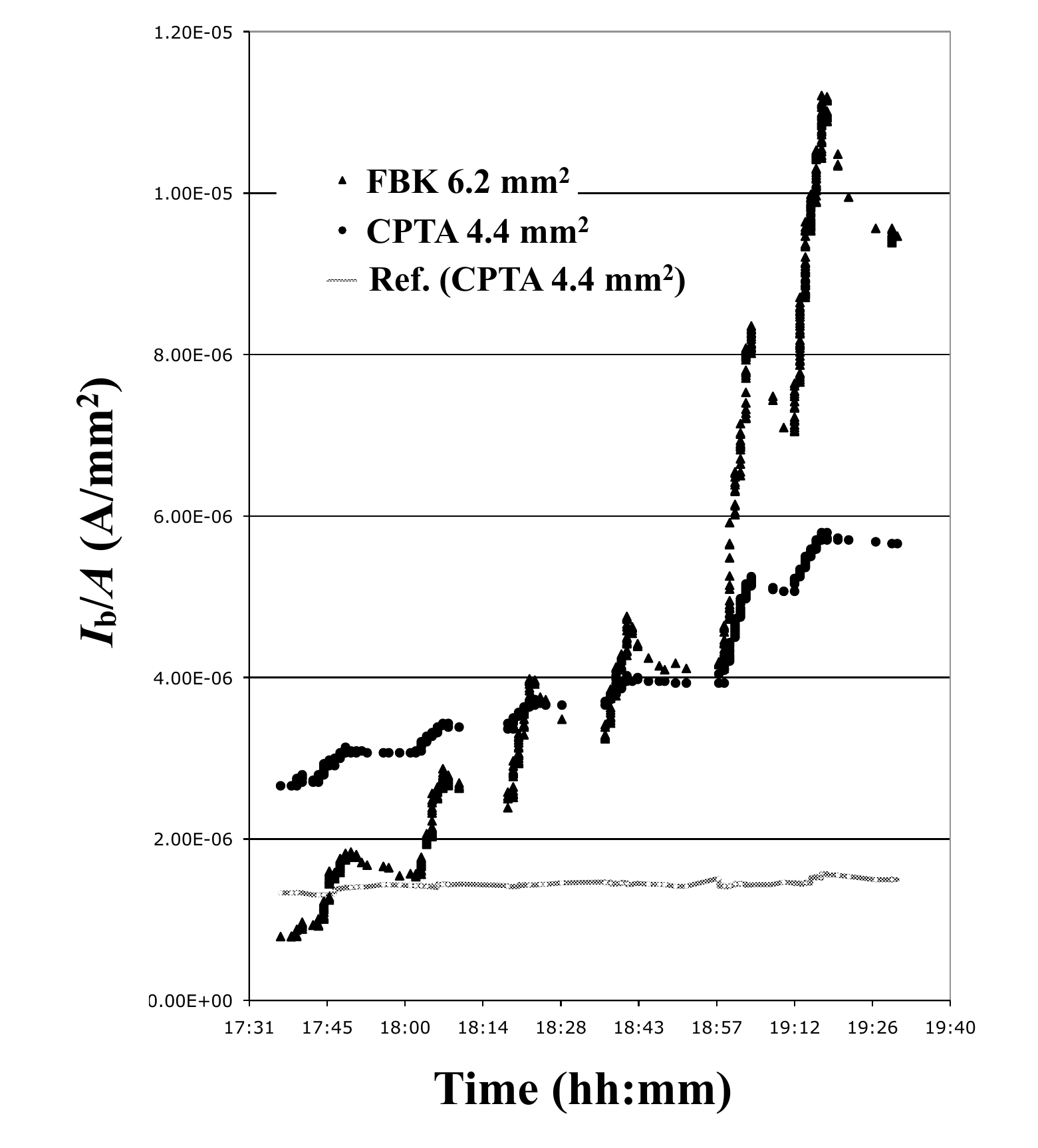}
  \caption{Leakage currents per area measured during irradiation for SiPMs on board 4:  CPTA 4.4 mm$^2$ reference diode (line), CPTA 4.4 mm$^2$ (circles), and FBK 6.2 mm$^2$  (triangles).  The plateaus correspond to partial fluences of $2.5\times 10^9$, $5\times 10^9$, $7.5\times 10^9$, $10^{10}$, $2\times 10^{10}$, and $3\times 10^{10}$ protons per cm$^2$. A drop in leakage current due to room-temperature annealing is visible after each step.}
\label{fig:leak3}
\end{figure}

Figure~\ref{fig:leak2} shows $I_{\rm b}/A$   {\it vs.} time for the larger-area SiPMs on board 3. 
The peaks correspond to partial fluences of $2.5\times 10^9$, $5\times 10^9$, $7.5\times 10^9$, and $10^{10}$ protons per cm$^2$. A drop in leakage current due to room-temperature annealing is visible after each step and is especially pronounced for the FBK 6.8 mm$^2$ SiPM.
Figure~\ref{fig:leak3} shows $I_{\rm b}/A$   {\it vs.} time for the  SiPMs on board 4. 
The peaks correspond to partial fluences of $2.5\times 10^9$, $5\times 10^9$, $7.5\times 10^9$, $10^{10}$, $2\times 10^{10}$, and $3\times 10^{10}$ protons per cm$^2$. The leakage currents for the single pixel readouts show a similar time structure with fluence and were in the 50-200 nA range.

\clearpage

\section {CPTA 4.4 mm$^2$ Reference SiPMs}

A CPTA 4.4 mm$^2$ SiPM was installed on each board and was not irradiated in order to serve as a reference signal to monitor the stability of the LED. The reference SiPMs were monitored before, during, and after the irradiation.  Table 2 shows the currents, gain, number of photoelectrons and signal stability for the reference SiPMs. The  data were taken at the same time that the indicated fluence was delivered to the other three SiPMs on each board. There was about an 8 h time difference between the first measurement  on board 1 and the last measurement on board 4. The measurements on 15 April 2008 were taken 135 days later. The values of $I_{\rm b}/A$ were stable for all the reference SiPMs. The calculated values of $MF$ from the mean and rms of the LED data were also stable at the few percent level. The values of $n_{\rm PE}/F$ were uniform to a few per cent, indicating that the light output of the LED  was relatively stable over this period. A change of the signal response ($S$) divided by the initial value ($S_0$) was observed to drift by 5-7\% over the measurement period. 

\begin{table}[h]
\begin{center}
\caption{Measured properties of the CPTA 4.4 mm$^2$ reference SiPMs: leakage current, gain, number of photoelectrons, and average response to the LED. The reference SiPMs were not irradiated.  The data were taken at the  time that the other SiPMs on the same board received the partial fluence indicated in column 2.}

\bigskip
\begin{tabular}[]{|c|c|c|c|c|c|}
\hline
{\it Board}   & {\it Time} & {$I_{\rm b}/A$  ($\mu$A/mm$^2$)} & {$MF$ (fC/PE)}  & { $n_{\rm PE}/F$} & {$S/S_0$} \\
\hline
\hline
{\rm 1} & {at zero} & {1.2} & {51} & {146} & {1} \\
\hline
{\rm 1} & {at $2.5\times 10^9$ cm$^{-2}$} & {1.1} & {50} & {148} & {1.00} \\
\hline
{\rm 1} & {at $5\times 10^9$ cm$^{-2}$} & {1.1} & {48} & {150} & {0.97} \\
\hline
{\rm 1} & {at $7.5\times 10^9$ cm$^{-2}$} & {1.1} & {50} & {144} & {0.96} \\
\hline
{\rm 1} & {at $10^{10}$ cm$^{-2}$} & {1.1} & {50} & {146} & {0.95} \\
\hline
{\rm 1} & {15Apr08} & {1.1} & {49} & {141} & {0.93} \\
\hline
\hline
{\rm 2} & {at zero} & {1.6} & {100} & {170} & {1} \\
\hline
{\rm 2} & {at $5\times 10^9$ cm$^{-2}$} & {1.6} & {100} & {170} & {0.98} \\
\hline
{\rm 2} & {at $10^{10}$ cm$^{-2}$} & {1.6} & {100} & {170} & {0.97} \\
\hline
{\rm 2} & {at $3\times 10^{10}$ cm$^{-2}$} & {1.8} & {100} & {160} & {0.93} \\
\hline
{\rm 2} & {15Apr08} & {1.5} & {110} & {150} & {1.00} \\
\hline
\hline
{\rm 3} & {zero} & {2.3} & {110} & {190} & {1} \\
\hline
{\rm 3} & {at $2.5\times 10^9$ cm$^{-2}$} & {2.1} & {100} & {200} & {0.94} \\
\hline
{\rm 3} & {at $5\times 10^{9}$ cm$^{-2}$} & {2.1} & {100} & {200} & {0.95} \\
\hline
{\rm 3} & {at $7.5\times 10^{10}$ cm$^{-2}$} & {2.2} & {100} & {190} & {0.95} \\
\hline
{\rm 3} & {at $10^{10}$ cm$^{-2}$} & {2.1} & {100} & {190} & {0.95} \\
\hline
{\rm 3} & {15Apr08} & {2.0} & {100} & {200} & {0.99} \\
\hline
\hline
{\rm 4} & {at zero} & {1.3} & {120} & {140} & {1} \\
\hline
{\rm 4} & {at $5\times 10^9$ cm$^{-2}$} & {1.4} & {110} & {150} & {0.98} \\
\hline
{\rm 4} & {at $10^{10}$ cm$^{-2}$} & {1.4} & {110} & {140} & {0.98} \\
\hline
{\rm 4} & {at $3\times 10^{10}$ cm$^{-2}$} & {1.5} & {110} & {140} & {0.93} \\
\hline
{\rm 4} & {15Apr08} & {1.4} & {20} & {140} & {0.99} \\
\hline
\end{tabular}
\end{center}
\label{tab:ref}
\end{table}

\section {CPTA 1.0 mm$^2$}
\begin{figure}[htbp]
  \centering
  \includegraphics[width=1\textwidth]{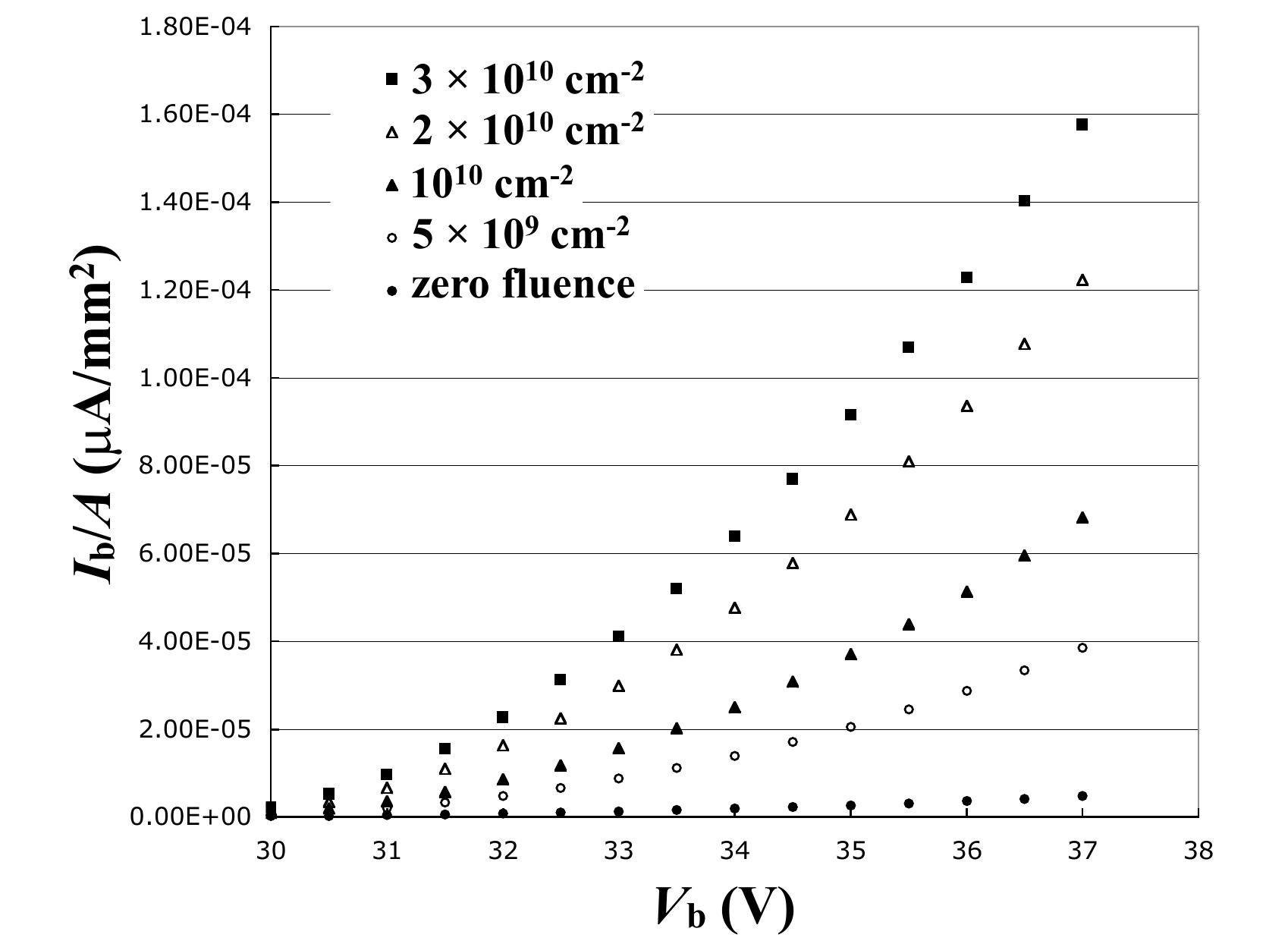}
  \caption{Leakage currents per mm$^2$ for CPTA1.0 mm$^2$ on board 2 as a function of bias voltage for varying proton fluence.}
\label{fig:leak_b5ch2}
\end{figure}

Measurements of $I_{\rm b}$ {\it vs.} $V_{\rm b}$ were taken before and after each partial fluence. Figure~\ref{fig:leak_b5ch2} shows $I_{\rm b}$ as a function of $V_{\rm b}$ for CPTA 1.0 mm$^2$ on board 2 for
fluences of zero, $5\times 10^9$, $10^{10}$, $2\times 10^{10}$, and  $3\times 10^{10}$ protons per cm$^2$. The shape of the $I_{\rm b}$ {\it vs.} $V_{\rm b}$ distributions indicate that the gain {\it vs.} voltage is relatively stable. 
A direct measurement of $MF$ from the mean and width of the response to the 
LED as a function of voltage before and after irradiation shows that the gain 
in the region of nominal voltage varies by about 100 fC/PE per V for CPTA 1.0 
mm$^2$ on board 1 and about 50  fC/PE per V for CPTA 1.0 mm$^2$ on board 2.
At a nominal operating voltage of $V_{\rm b}=34~\rm V$, the leakage current increases from 1.9 $\mu$A at zero fluence to 63.6 $\mu$A at $3\times 10^{10} \rm ~cm^{-2}$.
Similar $I_{\rm b}$ {\it vs.} $V_{\rm b}$ curves were observed for the other  CPTA 1.0 mm$^2$ on board 1, where the leakage current at nominal voltage (34 V) increased from  1.5 $\mu$A at zero fluence to 51 $\mu$A at $10^{10} \rm ~cm^{-2}$.

Table 3 shows the values of $I_{\rm b}$, $MF$, and $n_{\rm PE}/F$ as defined in section 2, as well as the change in signal $S$ in response to the LED divided by that at zero fluence ($S_0$). The values of $n_{\rm PE}$ are corrected for the measured deviation of the reference diode (see Table 2). The gain times excess noise factor is observed to decrease from 370 fC/PE to 300 fC/PE for board 1 and from 230 fC/PE to 180 fC/PE for board 2. 
At large bias currents, a drop in gain is expected due to a reduction in the bias voltage caused by a voltage drop across the 2 k$\Omega$ input resistor. 

\begin{table}[h]
\begin{center}
\caption{Measured properties of the CPTA 1.0 mm$^2$ SiPMs. The bias voltage was 34 V. }

\bigskip
\begin{tabular}[]{|c|c|c|c|c|c|}
\hline
{\it Board}   & {\it Fluence \rm ($\rm cm^{-2}$)} & {$I_{\rm b}/A$  ($\mu$A/mm$^2$)} & {$MF$ (fC/PE)}  & { $n_{\rm PE}/F$} & {$S/S_0$} \\
\hline
\hline
{\rm 1} & {zero} & {1.5 } & {370} & {46} & {1} \\
\hline
{\rm 1} & {$2.5\times 10^9$} & {16} & {330} & {50} & {0.96} \\
\hline
{\rm 1} & {$5\times 10^9$} & {23} & {340} & {48} & {0.96} \\
\hline
{\rm 1} & {$7.5\times 10^9$} & {39} & {320} & {50} & {0.92} \\
\hline
{\rm 1} & {$10^{10}$} & {51} & {300} & {51} & {0.88} \\
\hline
{\rm 1} & {15Apr08} & {30.4} & {310} & {51} & {0.94} \\
\hline
\hline
{\rm 2} & {zero} & {1.9 } & {230} & {44} & {1} \\
\hline
{\rm 2} & {$5\times 10^9$} & {13.9 } & {220} & {44} & {0.94} \\
\hline
{\rm 2} & {$10^{10}$} & {24.7} & {220} & {43} & {0.90} \\
\hline
{\rm 2} & {$3\times 10^{10}$} & {63.6 } & {180} & {42} & {0.75} \\
\hline
{\rm 2} & {15Apr08} & {35.8 } & {200} & {41} & {0.78} \\
\hline
\end{tabular}
\end{center}
\label{tab:b2ch2}
\end{table}

\begin{figure}[htbp]
  \centering
  \includegraphics[width=1\textwidth]{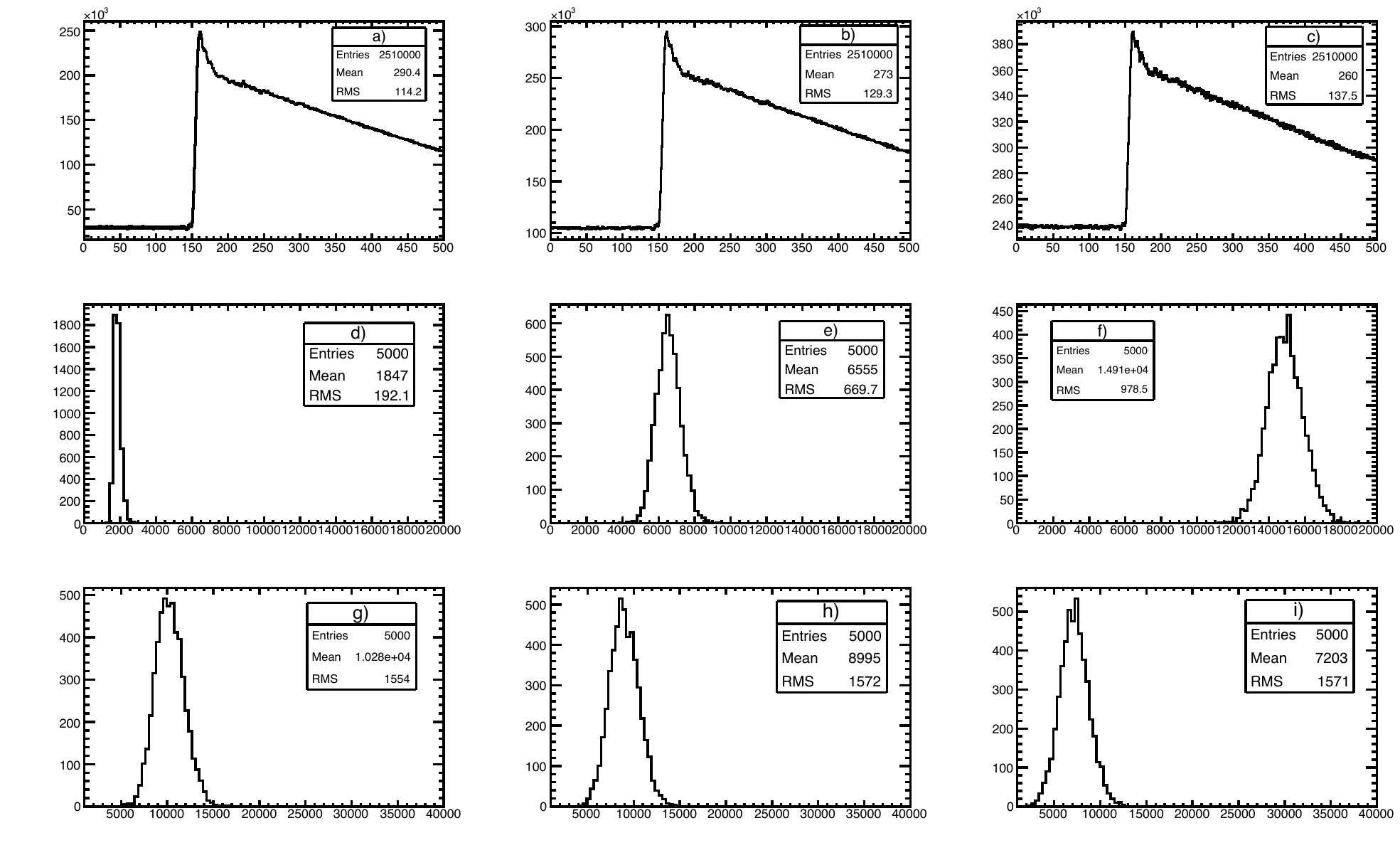}
  \caption{CPTA1.0 mm$^2$  at $V_{\rm b}= 34$ V on board 2:   
  pulse shape a) before irradiation, b) after $10^{10}~\rm cm^{-2}$, and c) after  $3\times 10^{10}~\rm cm^{-2}$; 
  noise distribution d) before irradiation, e) after $10^{10}~\rm cm^{-2}$, and f) after  $3\times 10^{10}~\rm cm^{-2}$; and
   signal distribution in response to LED g) before irradiation, h) after $10^{10}~\rm cm^{-2}$, and i) after  $3\times 10^{10}~\rm cm^{-2}$. 
}
\label{fig:b5ch2}
\end{figure}

The pulse shape in response to the LED was monitored in 500 2 ns time bins.
The pulse shape was observed to be stable at all fluences on both boards 1 and 2.
Figure~\ref{fig:b5ch2} shows the average pulse shape on baord 2 summed over 5000 events for
a) zero fluence, b) $10^{10} \rm ~cm^{-2}$, and c) $3\times 10^{10}\rm ~cm^{-2}$. The pulse shape of the CPTA 1.0 mm$^2$ was observed to have a long time-constant component due to a large value of quenching resistance. 

The pedestal was summed over 200 ns (bins 1-100 of fig.~\ref{fig:b5ch2} a),b), and c)) to get the noise distributions in fC shown in fig.~\ref{fig:b5ch2} for d) zero fluence, e) $10^{10} \rm ~cm^{-2}$, and f) $3\times 10^{10}\rm ~cm^{-2}$ for board 2. The rms noise increases from 192 fC at zero fluence to 670 fC at $10^{10} \rm ~cm^{-2}$ to 979 fC at $3\times 10^{10} \rm ~cm^{-2}$. The noise distribution for CPTA 1 mm$^2$ on board 1 was 277 fC at  zero fluence, increasing to 1205 fC at  $10^{10} \rm ~cm^{-2}$ for $V_{\rm b}= 34 \rm V$.

The pulse was summed over 200 ns (bins 151-250 of fig.~\ref{fig:b5ch2} a),b), and c)) and the pedestal was subtracted to get the signal distributions in fC shown in fig.~\ref{fig:b5ch2} for g) zero fluence, h) $10^{10} \rm ~cm^{-2}$, and i) $3\times 10^{10}\rm ~cm^{-2}$ for board 2. To calibrate out any instability of the LED, the change in signal was monitored relative to the CPTA 4.4 mm$^2$  reference SiPM on the same board. The variation of the signals from the reference SIPM varied by 5\% for board 1 and 7\% for board 2. The signal on the CPTA 1mm$^2$ SiPM on board 2, relative to  zero fluence and corrected for the reference diode signal, was observed to drop by 10\% at a fluence of $10^{10} \rm ~cm^{-2}$ and 25\% at $3\times 10^{10} \rm ~cm^{-2}$. Similarly, the signal for CPTA 1mm$^2$ on board 1 was 12\% lower at  a fluence of $10^{10} \rm ~cm^{-2}$.

\begin{figure}[htbp]
  \centering
  \includegraphics[width=1\textwidth]{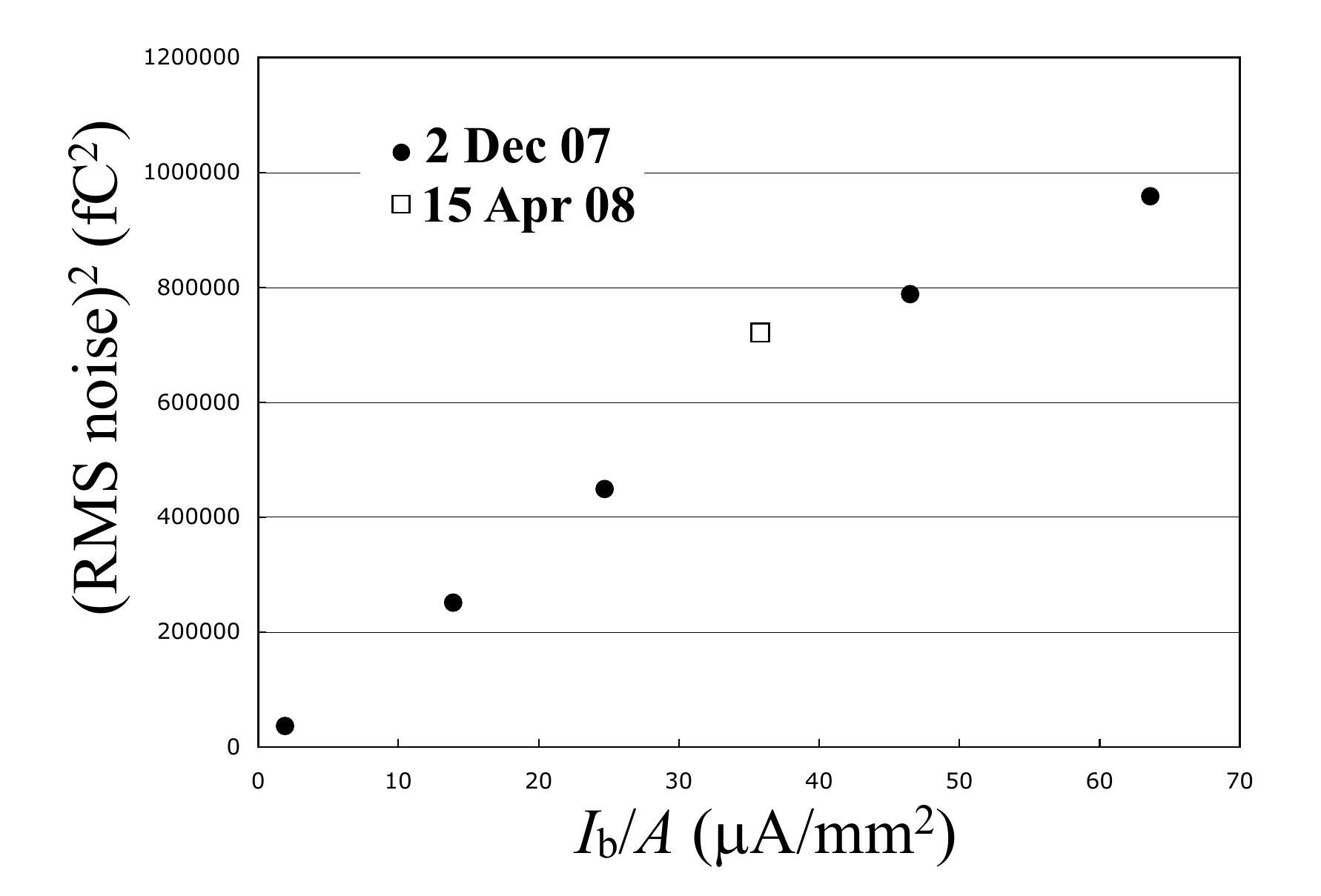}
  \caption{Pedestal rms noise squared {\it vs.} leakage current for CPTA 1.0 mm$^2$ on board 2, for data taken at the time of irradiation, 2 Dec 07 (solid circles) and after room temperature annealing on 15 Apr 08 (open square).}
\label{fig:b5_ch2_rms}
\end{figure}

Figure~\ref{fig:b5_ch2_rms} shows the square of the rms noise as a function of $I_{\rm b}/A$ for CPTA 1.0 mm$^2$ on board 2, for data taken immediately after the irradiation (2 Dec 07). The approximate linear dependance indicates that the increase in noise is due to an increase in rate of dark counts, {\it i.e.} that the leakage current is proportional to the square of the number of activated pixels. Detailed measurements were made after the irradiation as the SiPMs were allowed to anneal at room temperature. A substantal amount of annealing was observed. On 15 Apr 08, 135 days after the irradiation,  the dark current had dropped from 63.6 $\mu$A to 35.8 $\mu$A for  CPTA 1.0 mm$^2$ on board 2. The rms of the noise distribution on 15 Apr 08 was about 4\% larger than at the time of irradiation corresponding to the same leakage current as interpolated from the measurements at fluences of $10^{10}~\rm cm^{-2}$ and $2\times 10^{10}~\rm cm^{-2}$.

\section {HC 1.0 mm$^2$}

\begin{figure}[htbp]
  \centering
  \includegraphics[width=1\textwidth]{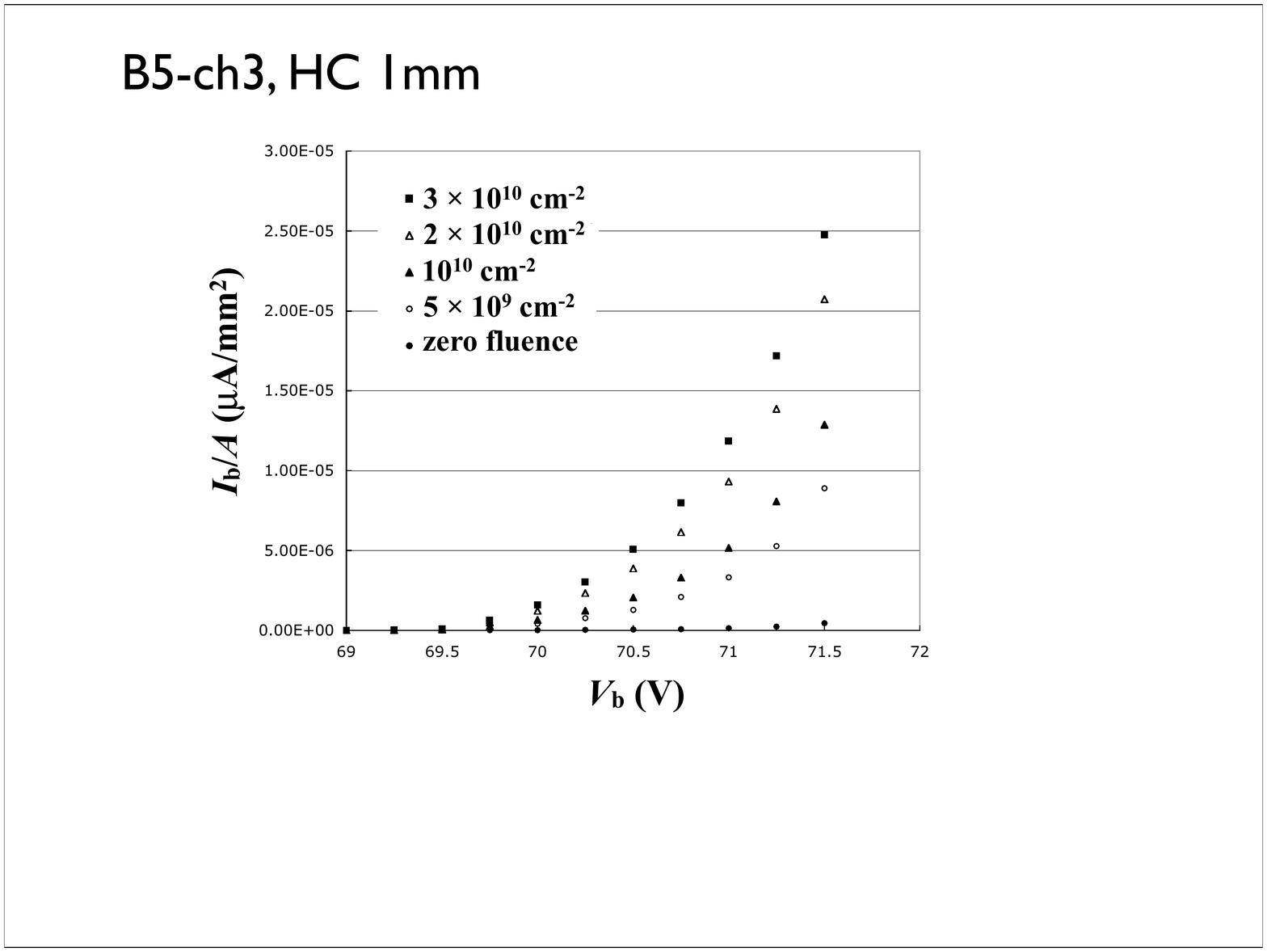}
  \caption{Leakage currents per mm$^2$ for HC 1.0 mm$^2$ on board 2 as a function of bias voltage for varying proton fluence.}
\label{fig:leak_b5ch3}
\end{figure}

Figure~\ref{fig:leak_b5ch3} shows $I_{\rm b}$ as a function of $V_{\rm b}$ for HC 1.0 mm$^2$ on board 2 for
fluences of zero, $5\times 10^9$, $10^{10}$, $2\times 10^{10}$, and  $3\times 10^{10}$ protons per cm$^2$. The shape of the $I_{\rm b}$ {\it vs.} $V_{\rm b}$ indicate that the gain {\it vs.} voltage is stable, although the turn-on with voltage is much steeper for the HC 1.0 mm$^2$ than for the CPTA 1.0 mm$^2$. 
A direct measurement of $MF$ as a function of voltage before and after irradiation 
shows that the gain in the region of nominal voltage varies by about 210 
fC/PE per V for HC 1.0 mm$^2$ on both board 1 and board 2.
At a nominal operating voltage of $V_{\rm b}=70.5~\rm V$, the leakage current increases from 0.05 $\mu$A at zero fluence to 5.6 $\mu$A at $3\times 10^{10} \rm ~cm^{-2}$.
Similar $I_{\rm b}$ {\it vs.} $V_{\rm b}$ curves were observed for the other  HC 1.0 mm$^2$ on board 1, where the leakage current at nominal voltage (70.5 V) increased from  0.1 $\mu$A at zero fluence to 2.5 $\mu$A at $10^{10} \rm ~cm^{-2}$.

Table 4 shows the values of $I_{\rm b}$, $MF$, and $n_{\rm PE}/F$ and $S/S_0$. The values of $n_{\rm PE}$ are again corrected for the measured deviation of the reference diode (see Table 2). The gain times excess noise factor is observed to decrease from 210 fC/PE to 180 fC/PE for board 1 and from 250 fC/PE to 210 fC/PE for board 2.
The HC SiPM is especially vulnerable to  
a drop in gain due to increased bias current because of its sharp turn-on.

\begin{table}[h]
\begin{center}
\caption{Measured properties of the HC 1.0 mm$^2$ SiPMs. The bias voltage was 70.5 V. }

\bigskip
\begin{tabular}[]{|c|c|c|c|c|c|}
\hline
{\it Board}   & {\it Fluence \rm ($\rm cm^{-2}$)} & {$I_{\rm b}/A$  ($\mu$A/mm$^2$)} & {$M F$ (fC/PE)}  & { $n_{\rm PE}/F$} & {$S/S_0$} \\
\hline
\hline
{\rm 1} & {zero} & {0.1 } & {250} & {48} & {1} \\
\hline
{\rm 1} & {$2.5\times 10^9$} & {0.9 } & {240} & {47} & {0.99}\\
\hline
{\rm 1} & {$5\times 10^9$} & {1.5} & {230} & {49} & {0.94} \\
\hline
{\rm 1} & {$7.5\times 10^9$} & {2.0} & {220} & {50} & {0.92} \\
\hline
{\rm 1} & {$10^{10}$} & {2.5} & {210} & {49} & {0.89} \\
\hline
{\rm 1} & {15Apr08} & {1.1} & {230} & {51} & {1.02} \\
\hline
\hline
{\rm 2} & {zero} & {0.05 } & {210} & {48} & {1}\\
\hline
{\rm 2} & {$5\times 10^9$} & {1.4 } & {210} & {47} & {0.99} \\
\hline
{\rm 2} & {$10^{10}$} & {2.1 } & {200} & {48} & {0.94} \\
\hline
{\rm 2} & {$3\times 10^{10}$} & {5.6 } & {180} & {47} & {0.85} \\
\hline
{\rm 2} & {15Apr08} & {2.3 } & {210} & {47} & {0.98}\\
\hline
\end{tabular}
\end{center}
\label{tab:det}
\end{table}

\begin{figure}[htbp]
  \centering
  \includegraphics[width=1\textwidth]{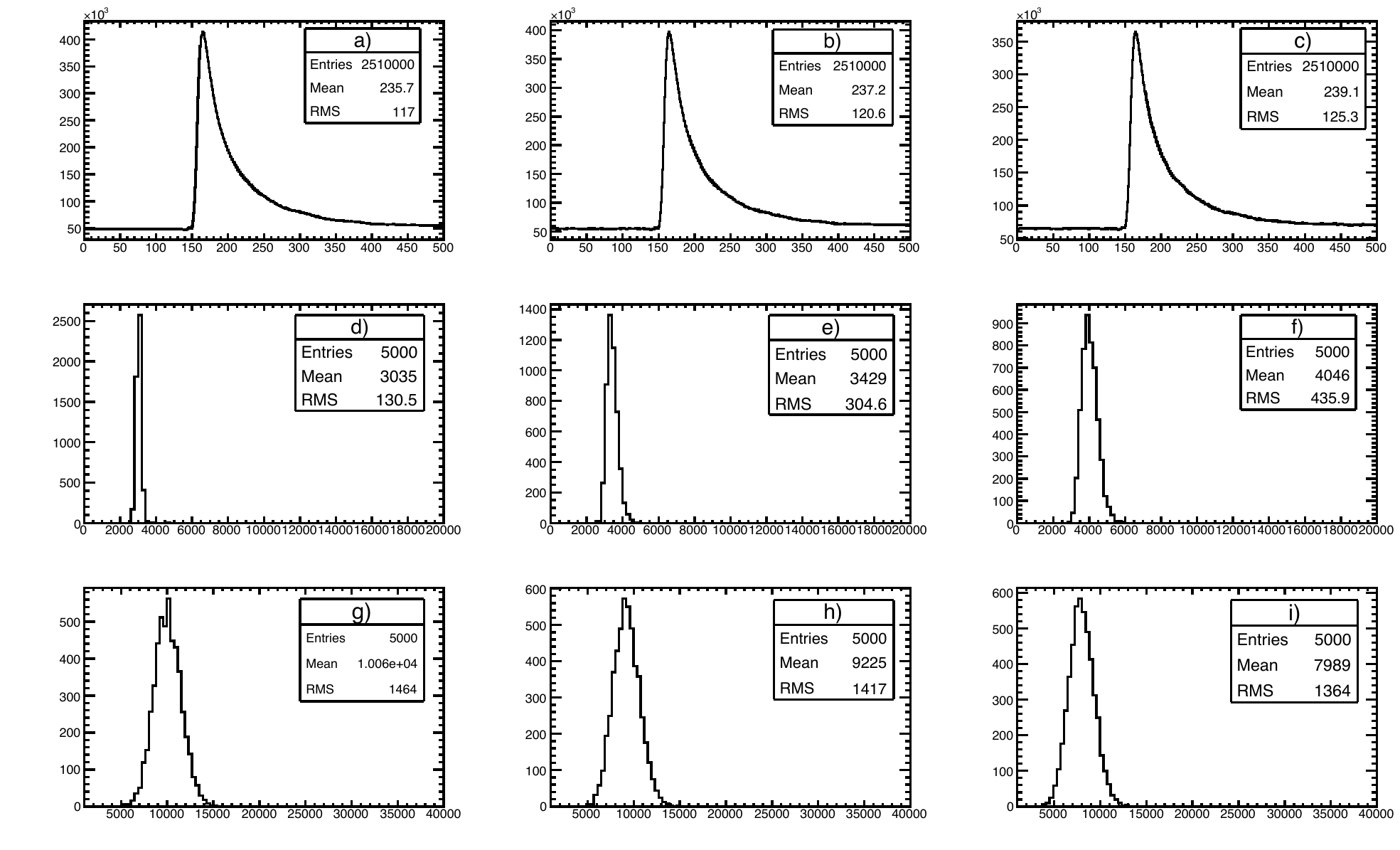}
  \caption{HC 1.0 mm$^2$  at $V_{\rm b}= 70.5$ V on board 2:   
  pulse shape a) before irradiation, b) after $10^{10}~\rm cm^{-2}$, and c) after  $3\times 10^{10}~\rm cm^{-2}$; 
  noise distribution d) before irradiation, e) after $10^{10}~\rm cm^{-2}$, and f) after  $3\times 10^{10}~\rm cm^{-2}$; and
   signal distribution in response to LED g) before irradiation, h) after $10^{10}~\rm cm^{-2}$, and i) after  $3\times 10^{10}~\rm cm^{-2}$. 
}
\label{fig:b5ch3}
\end{figure}

The pulse shape for 500 2 ns bins in response to the LED summed over 5000 events is shown in fig.~\ref{fig:b5ch3} for a) zero fluence, b) $10^{10} \rm ~cm^{-2}$, and c) $3\times 10^{10}\rm ~cm^{-2}$ for board 2. The pulse shape was observed to be stable at all fluences on both boards 1 and 2.

The pedestal was summed over 200 ns (bins 1-100 of fig.~\ref{fig:b5ch3} a),b), and c)) to get the noise distributions in fC shown in fig.~\ref{fig:b5ch3} for d) zero fluence, e) $10^{10} \rm ~cm^{-2}$, and f) $3\times 10^{10}\rm ~cm^{-2}$ for board 2. The rms noise increases from 131 fC at zero fluence to 305 fC at $10^{10} \rm ~cm^{-2}$ to 436 fC at $3\times 10^{10} \rm ~cm^{-2}$. 
The noise distribution for HC 1 mm$^2$ on board 1 was 126 fC at  zero fluence, increasing to 330 fC at  $10^{10} \rm ~cm^{-2}$ for $V_{\rm b}= 70.5 \rm V$.

The signal was summed over 200 ns (bins 151-250 of fig.~\ref{fig:b5ch3} a),b), and c)) and the noise was subtracted to get the signal distributions in fC shown in fig.~\ref{fig:b5ch3} for g) zero fluence, h) $10^{10} \rm ~cm^{-2}$, and i) $3\times 10^{10}\rm ~cm^{-2}$ for board 2. The signal on HC 1mm$^2$ on board 2, relative to  zero fluence and corrected for the reference diode signal, was observed to drop by 6\% at a fluence of $10^{10} \rm ~cm^{-2}$ and 15\% at $3\times 10^{10} \rm ~cm^{-2}$. Similarly, the signal for HC 1mm$^2$ on board 1 was 11\% lower at  a fluence of $10^{10} \rm ~cm^{-2}$.

\begin{figure}[htbp]
  \centering
  \includegraphics[width=1\textwidth]{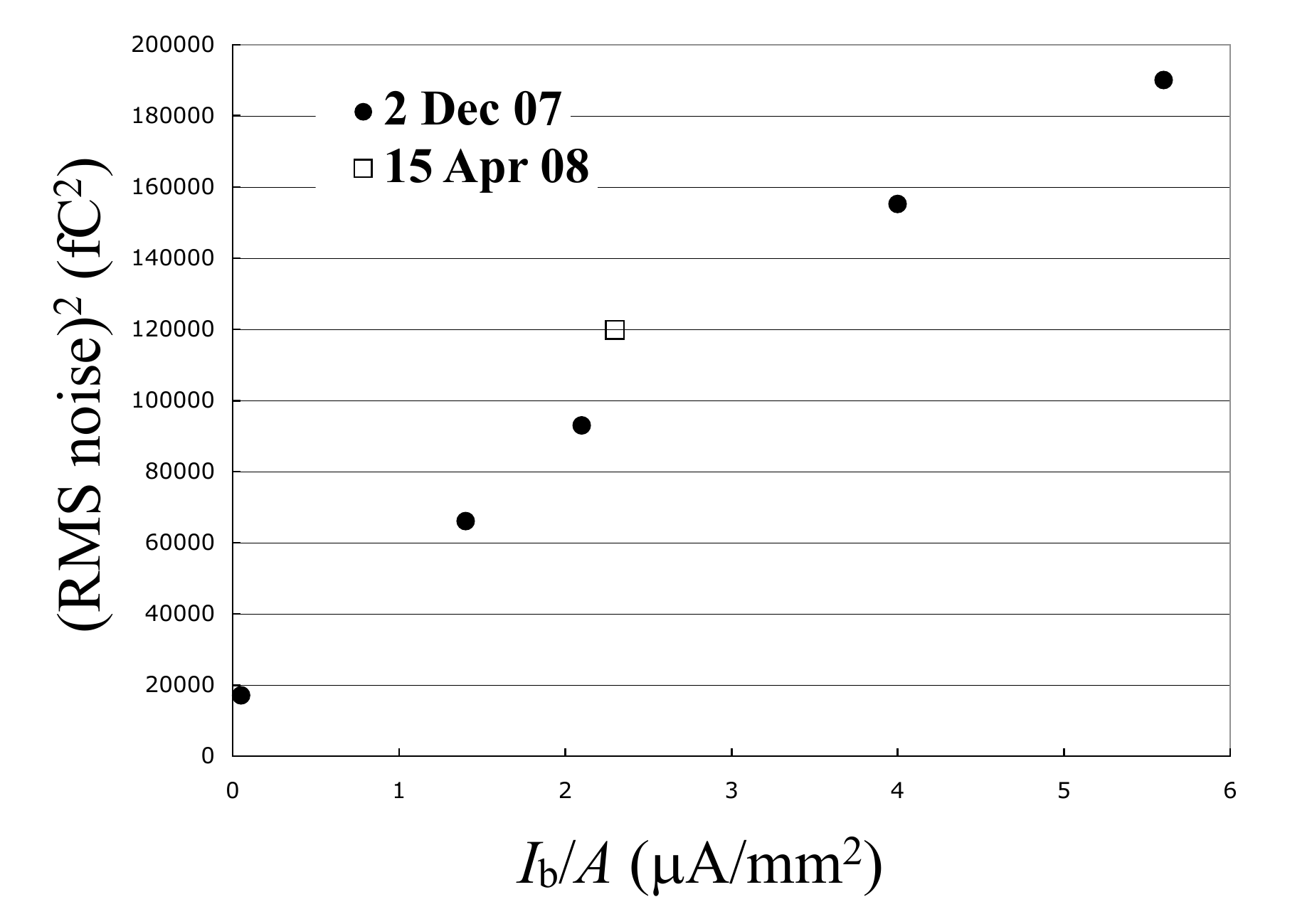}
  \caption{Pedestal rms noise squared {\it vs.} leakage current for HC 1.0 mm$^2$ on board 2, for data taken at the time of irradiation, 2 Dec 07 (solid circles) and after room temperature annealing on 15 Apr 08 (open square).}
\label{fig:b5_ch3_rms}
\end{figure}

Figure~\ref{fig:b5_ch3_rms} shows the square of the rms noise as a function of $I_{\rm b}/A$ for HC 1.0 mm$^2$ on board 2, for data taken immediately after the irradiation (2 Dec 07). The approximate linear dependance indicates that the increase in noise is due to an increase in rate of dark counts, {\it i.e.} that the leakage current is proportional to the square of the number of activated pixels. Detailed measurements were made after the irradiation as the SiPMs were allowed to anneal at room temperature. A substantial amount of annealing was observed. On 15 Apr 08, 135 days after the irradiation,  the dark current had dropped from 5.6 $\mu$A to 2.3 $\mu$A for  HC 1.0 mm$^2$ on board 2 and from 2.5 $\mu$A to 1.1 $\mu$A on board 1. The rms of the HC noise distribution on 15 Apr 08 was about 10\% larger than that at the time of irradiation corresponding to the same leakage current as interpolated from the measurements at fluences of $10^{10}~\rm cm^{-2}$ and $2\times 10^{10}~\rm cm^{-2}$.

\section{FBK 1.0 mm$^2$}
Figure~\ref{fig:leak_b5ch2} shows $I_{\rm b}$ as a funcion of $V_{\rm b}$ for FBK 1.0 mm$^2$ on board 2 for
fluences of zero, $5\times 10^9$, $10^{10}$, $2\times 10^{10}$, and  $3\times 10^{10}$ protons per cm$^2$. The shape of the $I_{\rm b}$ {\it vs.} $V_{\rm b}$ distributions indicate that the gain {\it vs.} voltage is again relatively stable. 
A direct measurement of $M F$ as a function of voltage before and after 
irradiation shows that the gain in the region of nominal voltage varies by  
about 170 fC/PE per V for FBK 1.0 mm$^2$ on board 1 and 110 fC/PE per V board 2.
At a nominal operating voltage of $V_{\rm b}=33.5~\rm V$, the leakage current increases from 1.6 $\mu$A at zero fluence to 20.8 $\mu$A at $3\times 10^{10} \rm ~cm^{-2}$.
Similar $I_{\rm b}$ {\it vs.} $V_{\rm b}$ curves were observed for the other  FBK 1.0 mm$^2$ on board 1, where the leakage current at nominal voltage (33.5 V) increased from  1.6 $\mu$A at zero fluence to 6.5 $\mu$A at $10^{10} \rm ~cm^{-2}$.

\begin{figure}[htbp]
  \centering
  \includegraphics[width=1\textwidth]{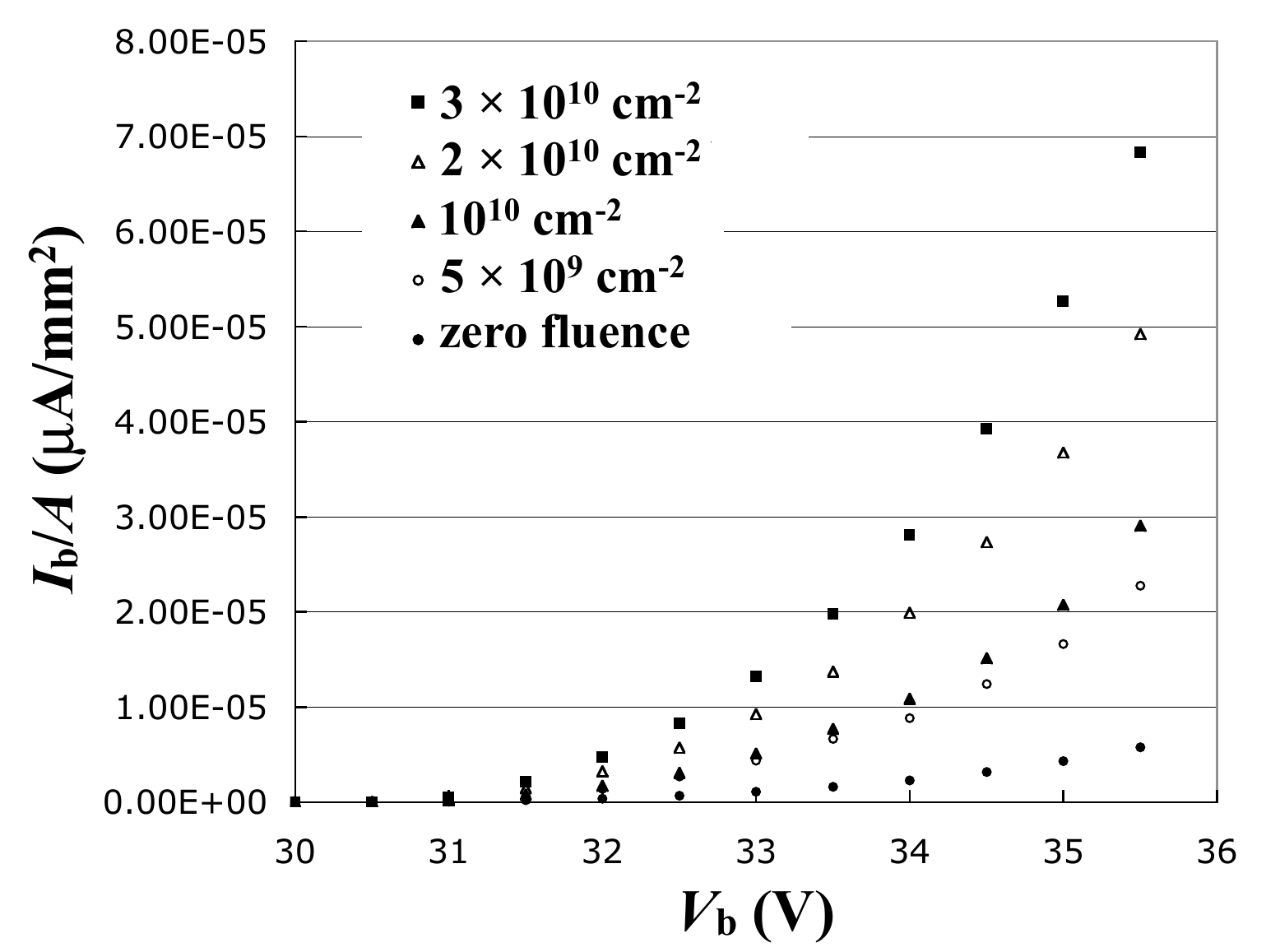}
  \caption{Leakage currents per mm$^2$ for FBK 1.0 mm$^2$ on board 2 as a function of bias voltage for varying proton fluence.}
\label{fig:leak_b5ch4}
\end{figure}

Table 5 shows the values of $I_{\rm b}$, $MF$, and $n_{\rm PE}/F$ and $S/S_0$. The values of $n_{\rm PE}$ are again corrected for the measured deviation of the reference diode (see Table 2). The gain times excess noise factor is observed to be stable in the range 430-450 fC/PE.

\begin{table}[h]
\begin{center}
\caption{Measured properties of the FBK 1.0 mm$^2$ SiPMs at $V_{\rm b}=33.5 \rm ~V$. The data of 15Apr08 were taken after after 135 days of room temperature annealing.}

\bigskip
\begin{tabular}[]{|c|c|c|c|c|c|}
\hline
{\it Board}   & {\it Fluence \rm ($\rm cm^{-2}$)} & {$I_{\rm b}/A$  ($\mu$A/mm$^2$)} & {$MF$ (fC/PE)}  & { $n_{\rm PE}/F$} & {$S/S_0$} \\
\hline
\hline
{\rm 1} & {zero} & {1.6 } & {430} & {39} & {1} \\
\hline
{\rm 1} & {$2.5\times 10^9$} & {2.4 } & {440} & {38} & {1.01} \\
\hline
{\rm 1} & {$5\times 10^9$} & {4.9} & {440} & {37} & {1.00} \\
\hline
{\rm 1} & {$7.5\times 10^9$} & {5.5} & {450} & {37} & {1.00} \\
\hline
{\rm 1} & {$10^{10}$} & {6.5} & {450} & {37} & {1.00}  \\
\hline
{\rm 1} & {15Apr08} & {3.9} & {470} & {38} & {1.08} \\
\hline
\hline
{\rm 2} & {zero} & {1.6 } & {460} & {35} &{1} \\
\hline
{\rm 2} & {$5\times 10^9$} & {5.4 } & {460} & {35} & {0.98}\\
\hline
{\rm 2} & {$10^{10}$} & {7.8 } & {480} & {33} & {0.96} \\
\hline
{\rm 2} & {$3\times 10^{10}$} & {20.8 } & {420} & {37} & {0.89} \\
\hline
{\rm 2} & {15Apr08} & {10.7 } & {450} & {33} & {0.92} \\
\hline
\end{tabular}
\end{center}
\label{tab:det}
\end{table}

\begin{figure}[htbp]
  \centering
  \includegraphics[width=1\textwidth]{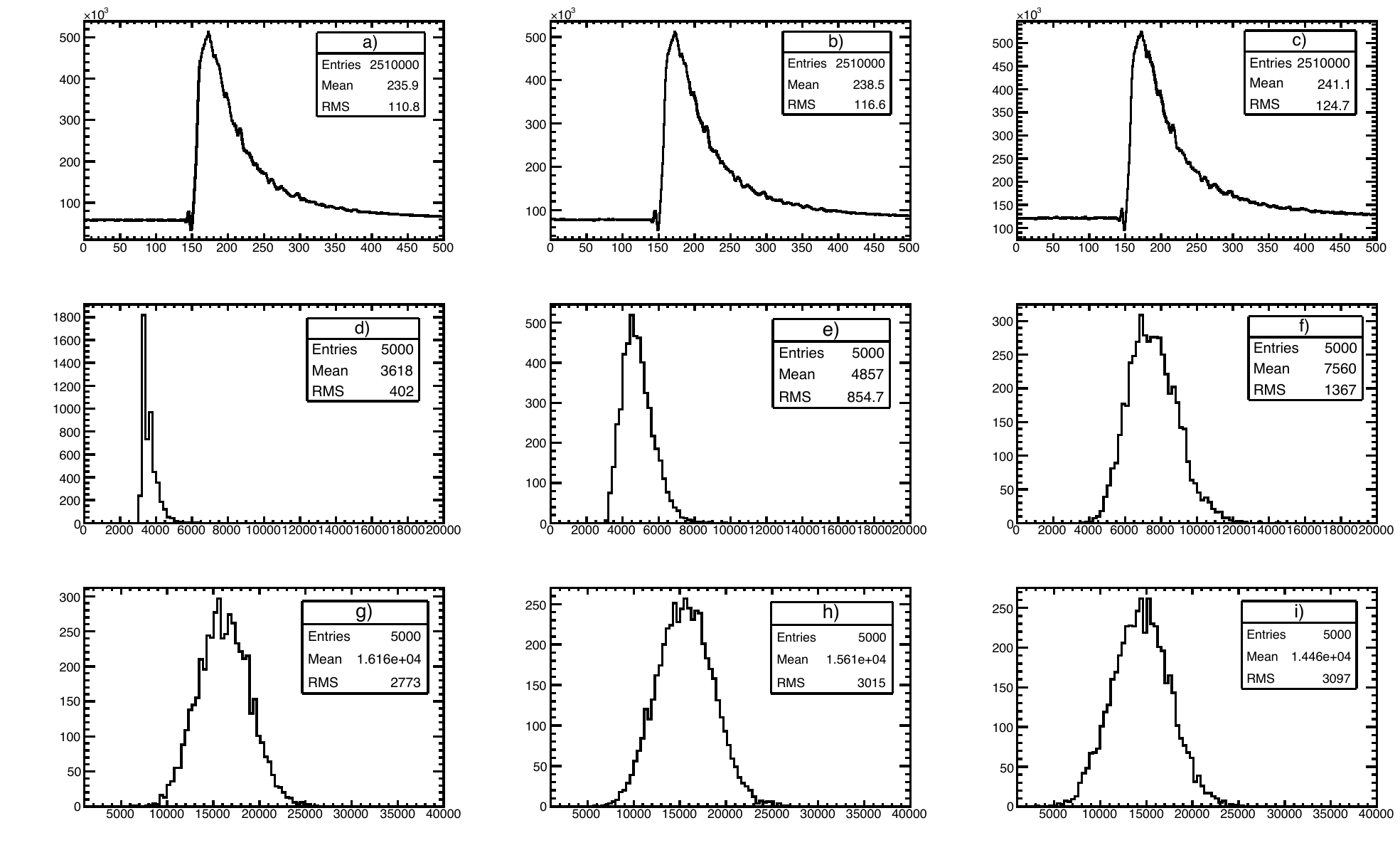}
  \caption{FBK 1.0 mm$^2$  at $V_{\rm b}= 33.5$ V on board 2:   
  pulse shape a) before irradiation, b) after $10^{10}~\rm cm^{-2}$, and c) after  $3\times 10^{10}~\rm cm^{-2}$; 
  noise distribution d) before irradiation, e) after $10^{10}~\rm cm^{-2}$, and f) after  $3\times 10^{10}~\rm cm^{-2}$; and
   signal distribution in response to LED g) before irradiation, h) after $10^{10}~\rm cm^{-2}$, and i) after  $3\times 10^{10}~\rm cm^{-2}$. 
}
\label{fig:b5ch4}
\end{figure}

The pulse shape for 500 2 ns bins in response to the LED summed over 5000 events is shown in fig.~\ref{fig:b5ch4} for a) zero fluence, b) $10^{10} \rm ~cm^{-2}$, and c) $3\times 10^{10}\rm ~cm^{-2}$ for board 2. The pulse shape was observed to be stable at all fluences on both boards 1 and 2.

The pedestal was summed over 200 ns (bins 1-100 of fig.~\ref{fig:b5ch4} a),b), and c)) to get the noise distributions in fC shown in fig.~\ref{fig:b5ch4} for d) zero fluence, e) $10^{10} \rm ~cm^{-2}$, and f) $3\times 10^{10}\rm ~cm^{-2}$ for board 2. The rms noise increases from 402  fC at zero fluence to 855 fC at $10^{10} \rm ~cm^{-2}$ to 1367 fC at $3\times 10^{10} \rm ~cm^{-2}$. 
The noise distribution for FBK 1 mm$^2$ on board 1 was 404 fC at  zero fluence, increasing to 720 fC at  $10^{10} \rm ~cm^{-2}$ for $V_{\rm b}= 33.5 \rm V$.

The signal was summed over 200 ns (bins 151-250 of fig.~\ref{fig:b5ch4} a),b), and c)) and the noise was subtracted to get the signal distributions in fC shown in fig.~\ref{fig:b5ch4} for g) zero fluence, h) $10^{10} \rm ~cm^{-2}$, and i) $3\times 10^{10}\rm ~cm^{-2}$ for board 2. The signal on FBK 1mm$^2$ on board 2, relative to  zero fluence and corrected for the reference diode signal, was observed to drop by 4\% at a fluence of $10^{10} \rm ~cm^{-2}$ and 11\% at $3\times 10^{10} \rm ~cm^{-2}$. The signal for FBK 1mm$^2$ on board 1 was observed not to change at  a fluence of $10^{10} \rm ~cm^{-2}$.

\begin{figure}[htbp]
  \centering
  \includegraphics[width=1\textwidth]{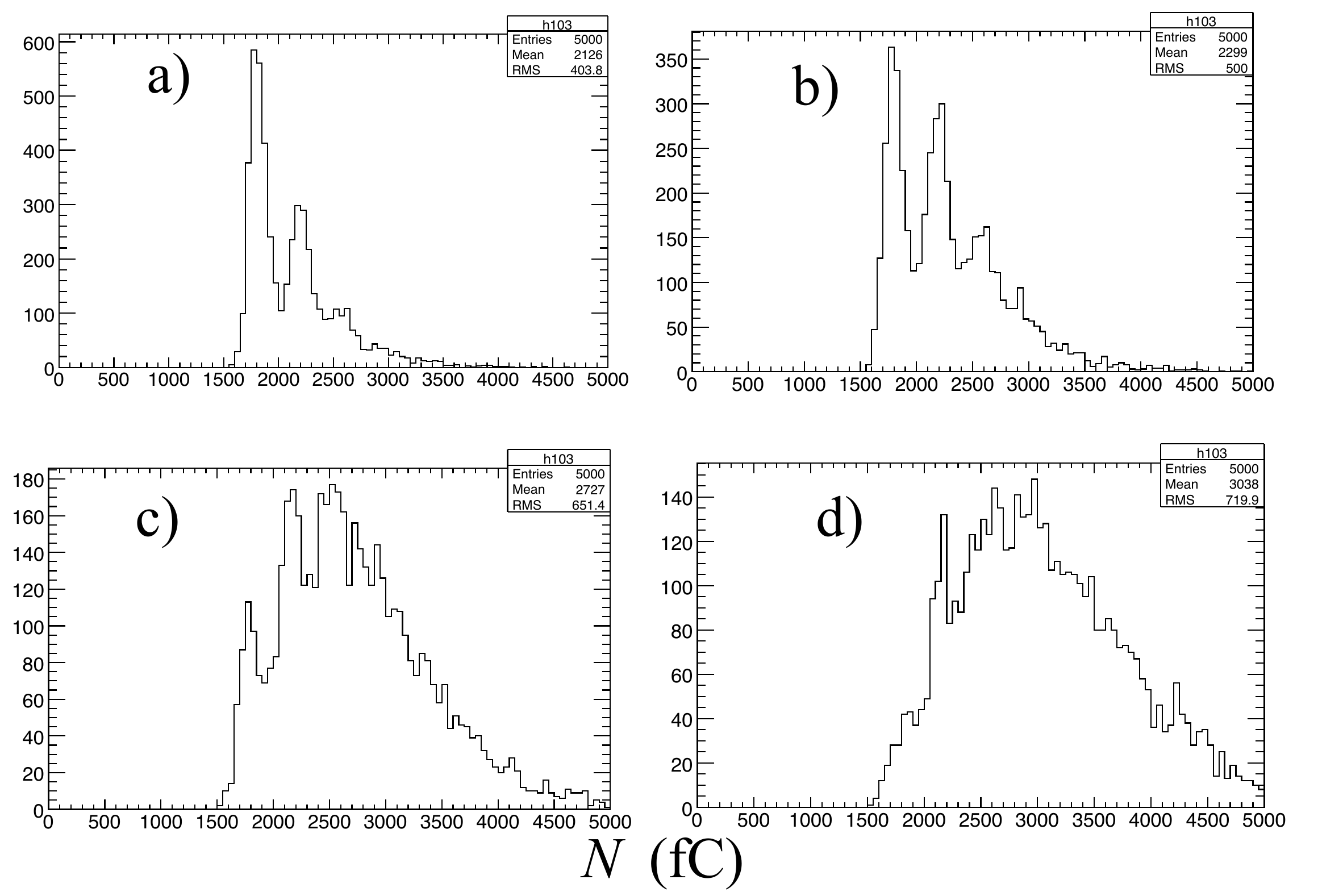}
  \caption{Noise distribution ($N$)  for FBK 1.0 mm$^2$ on board 1 a) before irradiation b) after a fluence of $2.5\times 10^9$ cm$^{-2}$, c) after a fluence of $5\times 10^9$ cm$^{-2}$, and d) after a fluence of $10^{10}$ cm$^{-2}$.}
\label{fig:single_pe}
\end{figure}

The noise ($N$) distribution for FBK 1.0 mm$^2$ shows a clear separation for zero and single PE.
Figure~\ref{fig:single_pe} shows the distribution of $N$ for FBK 1.0 mm$^2$ on board 1 a) before irradiation,
b) after a fluence of $2.5\times 10^9$ cm$^{-2}$, c) after a fluence of $5\times 10^9$ cm$^{-2}$, and d) after a fluence of $10^{10}$ cm$^{-2}$. Fits to the zero and single PE peaks give a gain of 370 fC/PE. Comparison of the value of MF as determined from the mean and width of the LED response (see Table 5) to the single PE peak gives $F = 1.2$, in agreement with previous measurements~\cite{sipm1}. Locations of the zero and single PE peaks do not change with irradiation, providing additional evidence that the gain is stable. A similar single PE peak and gain is found for FBK 1.0 mm$^2$ on board 2.

Figure~\ref{fig:leak_b5ch4} shows the square of the rms noise as a function of $I_{\rm b}/A$ for FBK 1.0 mm$^2$ on board 2, for data taken immediately after the irradiation (2 Dec 07). The linear dependance indicates that the increase in noise is due to an increase in rate of dark counts, {\it i.e.} that the leakage current is proportional to the square of the number of activated pixels.  On 15 Apr 08, 135 days after the irradiation,  the dark current had dropped from 5.6 $\mu$A to 2.3 $\mu$A for  FBK 1.0 mm$^2$ on board 2 and from 2.5 $\mu$A to 1.1 $\mu$A on board 1 indicating a substantial amount of annealing at room temperature. The rms of the noise distribution on 15 Apr 08 was about 2\% larger than at the time of irradiation corresponding to the same leakage current as interpolated from the measurements at fluences of $10^{10}~\rm cm^{-2}$ and $2\times 10^{10}~\rm cm^{-2}$.

\begin{figure}[htbp]
  \centering
  \includegraphics[width=1\textwidth]{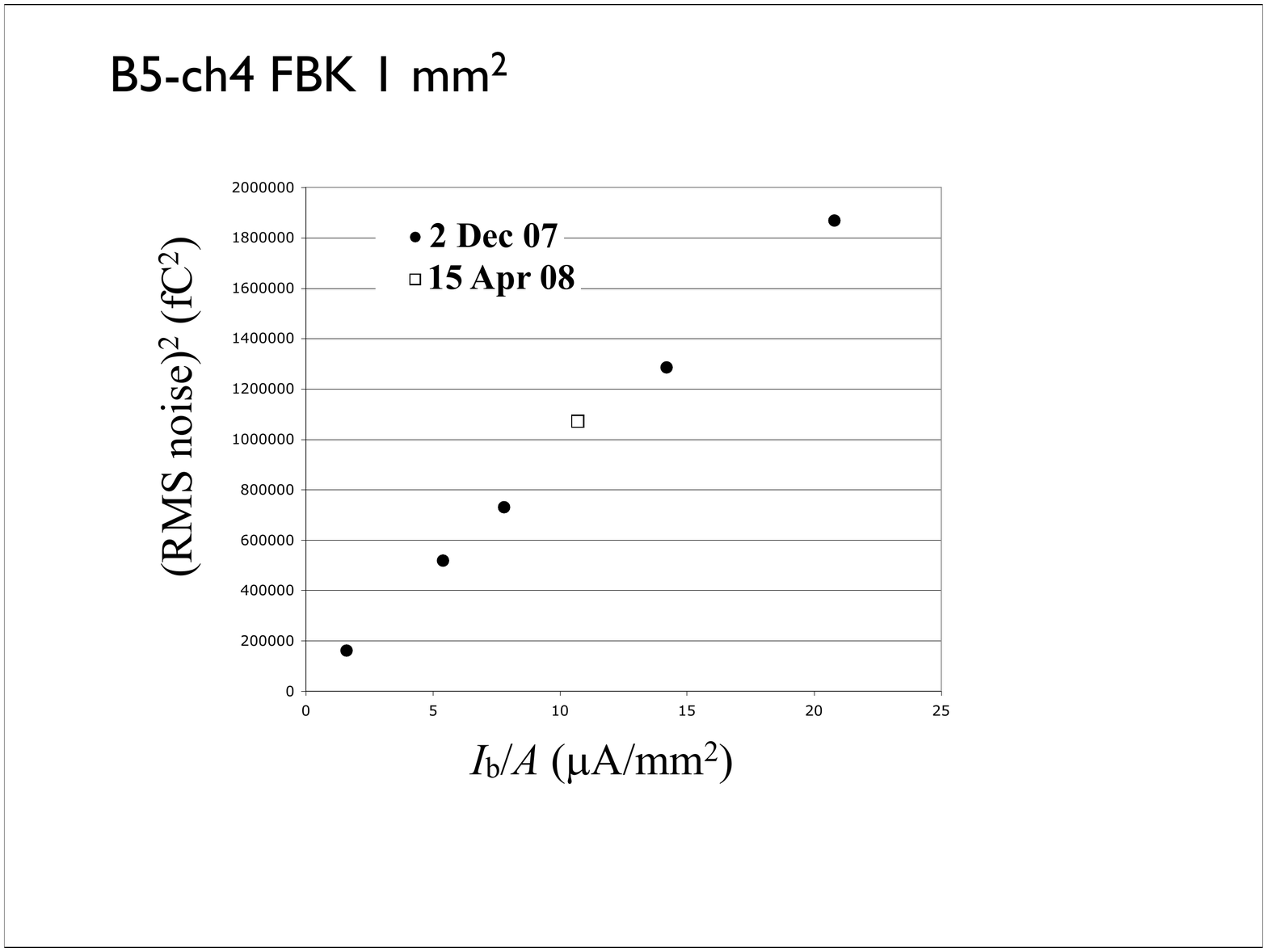}
  \caption{Pedestal rms noise squared {\it vs.} leakage current for FBK 1.0 mm$^2$ on board 2, for data taken at the time of irradiation, 2 Dec 07 (solid circles) and after room temperature annealing on 15 Apr 08 (open square).}
\label{fig:leak_b5ch4}
\end{figure}

\section {CPTA 4.4 mm$^2$}
\begin{figure}[htbp]
  \centering
  \includegraphics[width=1\textwidth]{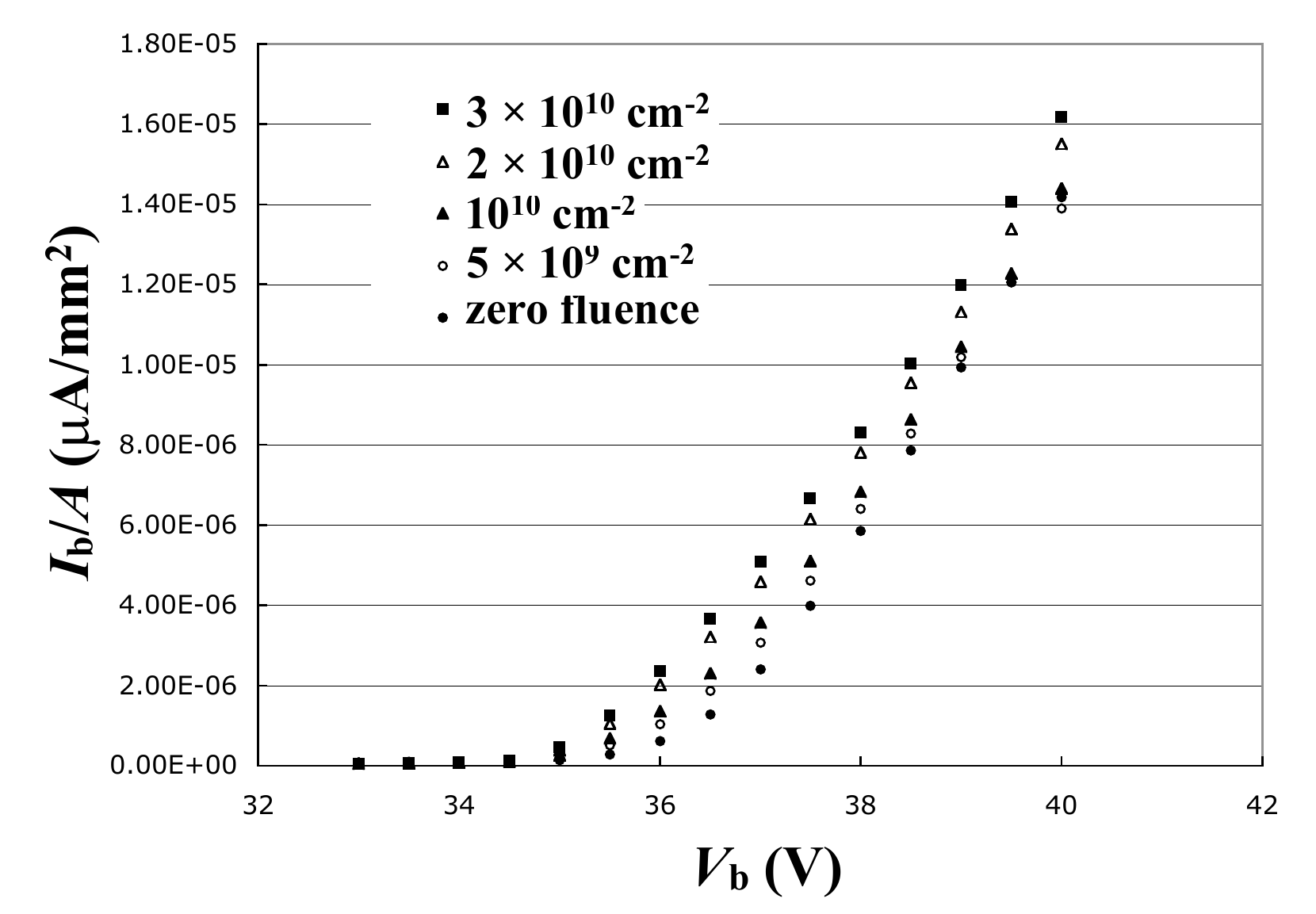}
  \caption{Leakage currents per mm$^2$ for CPTA 4.4 mm$^2$ on board 4 as a function of bias voltage for varying proton fluence.}
\label{fig:leak_b7ch2}
\end{figure}

Figure~\ref{fig:leak_b7ch2} shows $I_{\rm b}$ as a function of $V_{\rm b}$ for CPTA 4.4 mm$^2$ on board 4 for
fluences of zero, $5\times 10^9$, $10^{10}$, $2\times 10^{10}$, and  $3\times 10^{10}$ protons per cm$^2$. The shape of the $I_{\rm b}$ {\it vs.} $V_{\rm b}$ indicate that the gain {\it vs.} voltage is stable. 
A direct 
measurement of $M F$ as a function of voltage before and after irradiation shows 
that the gain in the region of nominal voltage varies by about 19 fC/PE per V for 
CPTA 4.4 mm$^2$ on board 3 and 12 fC/PE per V for board 4.
At a nominal operating voltage of $V_{\rm b}=37~\rm V$, the leakage current increases from 2.6 $\mu$A/mm$^2$ at zero fluence to 5.7 $\mu$A/mm$^2$ at $3\times 10^{10} \rm ~cm^{-2}$ for board 4.
Similar $I_{\rm b}$ {\it vs.} $V_{\rm b}$ curves were observed for the other  CPTA 4.4 mm$^2$ on board 3, where the leakage current at nominal voltage (37 V) increased from  1.8 $\mu$A/mm$^2$ at zero fluence to 4.1 $\mu$A/mm$^2$ at $10^{10} \rm ~cm^{-2}$.

Table 6 shows the values of $I_{\rm b}$, $MF$, and $n_{\rm PE}/F$ and $S/S_0$. The values of $n_{\rm PE}$ are again corrected for the measured deviation of the reference diode (see Table 2). The gain times excess noise factor is observed to decrease from 66 fC/PE to 61 fC/PE for board 1 and from 53 fC/PE to 43 fC/PE for board 2.

\begin{table}[h]
\begin{center}
\caption{Measured properties of the CPTA 4.4 mm$^2$ SiPMs. The bias voltage was 37 V. }
\bigskip
\begin{tabular}[]{|c|c|c|c|c|c|}
\hline
{\it Board}   & {\it Fluence \rm ($\rm cm^{-2}$)} & {$I_{\rm b}/A$  ($\mu$A/mm$^2$)} & {$M F$ (fC/PE)}  & { $n_{\rm PE}/F$} & {$S/S_0$} \\
\hline
\hline
{\rm 3} & {zero} & {1.8 } & {66} & {180} & {1} \\
\hline
{\rm 3} & {$2.5\times 10^9$} & {2.6} & {62} & {180} & {0.93} \\
\hline
{\rm 3} & {$5\times 10^9$} & {3.3} & {62} & {170} & {0.87} \\
\hline
{\rm 3} & {$7.5\times 10^9$} & {3.8} & {60} & {160} & {0.81} \\
\hline
{\rm 3} & {$10^{10}$} & {4.1} & {59} & {150} & {0.76} \\
\hline
{\rm 3} & {15Apr08} & {3.0} & {61} & {150} & {0.76}\\
\hline
\hline
{\rm 4} & {zero} & {2.6} & {53} & {152} &{1} \\
\hline
{\rm 4} & {$2.5\times 10^9$} & {3.0} & {52} & {145}& {0.93} \\
\hline
{\rm 4} & {$5\times 10^9$} & {3.4} & {54} & {133} & {0.89} \\
\hline
{\rm 4} & {$7.5\times 10^9$} & {3.6 } & {51} & {134} & {0.84} \\
\hline
{\rm 4} & {$10^{10}$} & {3.9} & {52} & {123} & {0.80} \\
\hline
{\rm 4} & {$3\times 10^{10}$} & {5.7} & {43} & {95} & {0.51} \\
\hline
{\rm 4} & {15Apr08} & {3.9} & {45} & {102} & {0.57} \\
\hline
\end{tabular}
\end{center}
\label{tab:det}
\end{table}

\begin{figure}[htbp]
  \centering
  \includegraphics[width=1\textwidth]{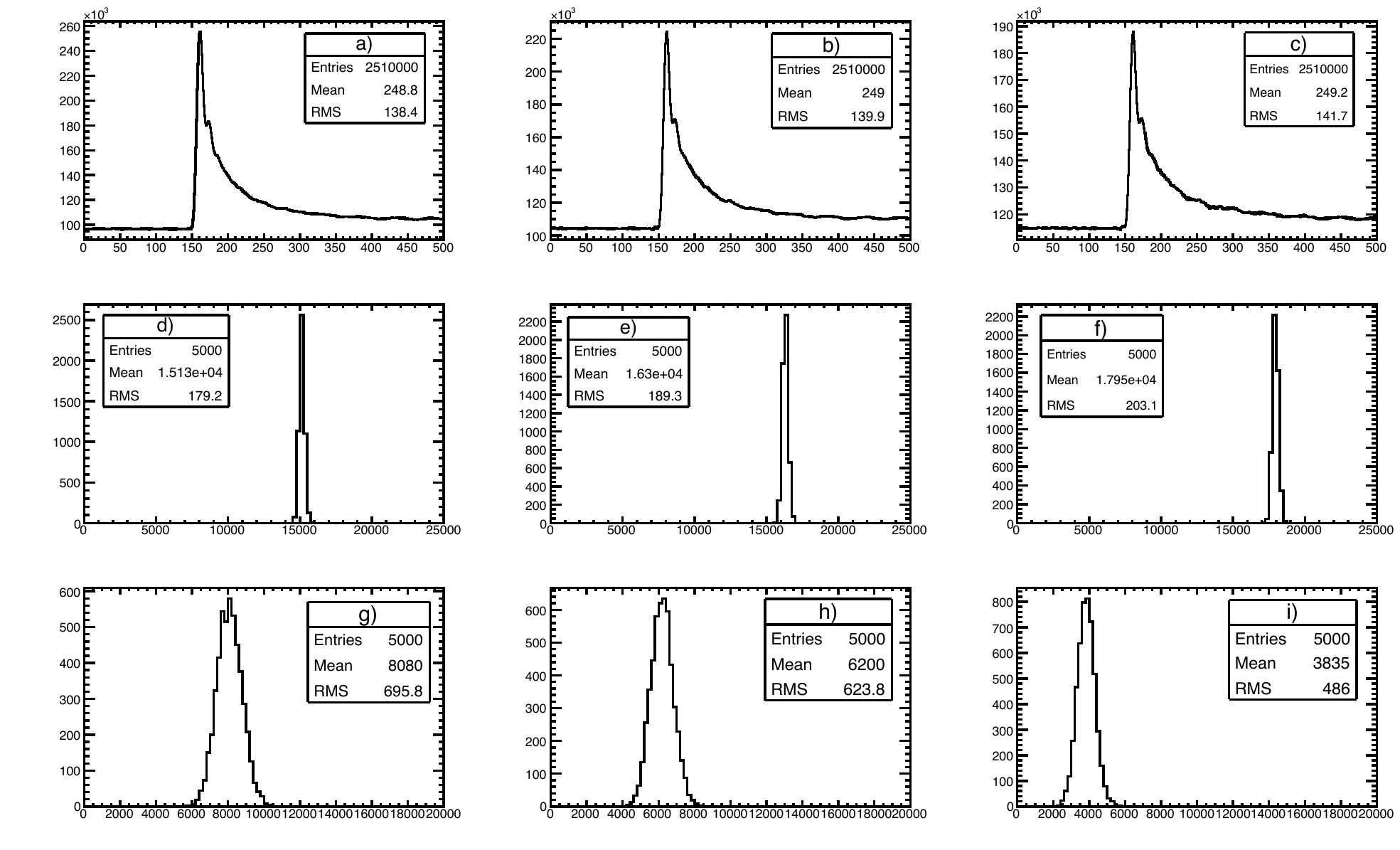}
  \caption{CPTA 4.4 mm$^2$  at $V_{\rm b}= 37$ V on board 4:   
  pulse shape a) before irradiation, b) after $10^{10}~\rm cm^{-2}$, and c) after  $3\times 10^{10}~\rm cm^{-2}$; 
  noise distribution d) before irradiation, e) after $10^{10}~\rm cm^{-2}$, and f) after  $3\times 10^{10}~\rm cm^{-2}$; and
   signal distribution in response to LED g) before irradiation, h) after $10^{10}~\rm cm^{-2}$, and i) after  $3\times 10^{10}~\rm cm^{-2}$. 
}
\label{fig:b7ch2}
\end{figure}

The pulse shape for 500 2 ns bins in response to the LED summed over 5000 events is shown in fig.~\ref{fig:b7ch2} for a) zero fluence, b) $10^{10} \rm ~cm^{-2}$, and c) $3\times 10^{10}\rm ~cm^{-2}$ for board 4. The pulse shape was observed to be stable at all fluences on both boards 3 and 4.

The pedestal was summed over 200 ns (bins 1-100 of fig.~\ref{fig:b7ch2} a),b), and c)) to get the noise distributions in fC shown in fig.~\ref{fig:b7ch2} for d) zero fluence, e) $10^{10} \rm ~cm^{-2}$, and f) $3\times 10^{10}\rm ~cm^{-2}$ for board 4. The rms noise increases from 179  fC at zero fluence to 189 fC at $10^{10} \rm ~cm^{-2}$ to 203 fC at $3\times 10^{10} \rm ~cm^{-2}$. 

The signal was summed over 200 ns (bins 151-250 of fig.~\ref{fig:b7ch2} a),b), and c)) and the noise was subtracted to get the signal distributions in fC shown in fig.~\ref{fig:b7ch2} for g) zero fluence, h) $10^{10} \rm ~cm^{-2}$, and i) $3\times 10^{10}\rm ~cm^{-2}$ for board 4. The signal on CPTA 4.4 mm$^2$ on board 4, relative to  zero fluence and corrected for the reference diode signal, was observed to drop by 20\% at a fluence of $10^{10} \rm ~cm^{-2}$ and 49\% at $3\times 10^{10} \rm ~cm^{-2}$. Similarly, the signal for CPTA 4.4 mm$^2$ on board 3 was 24\% lower at  a fluence of $10^{10} \rm ~cm^{-2}$.
The drop in signal is due in part to a substantial drop in the number of photoelectrons (38\% drop for board 4). This is due to the fact that the noise has increased to the point where a significant number of pixels are not available to respond to the LED light.
This saturation is also seen in fig.~\ref{fig:leak3} where the slope of the leakage current {\it vs.} time shows a flatttening during the last irradiation from $2\times 10^{10}$ to $3\times 10^{10}$ protons per cm$^2$.

\begin{figure}[htbp]
  \centering
  \includegraphics[width=1\textwidth]{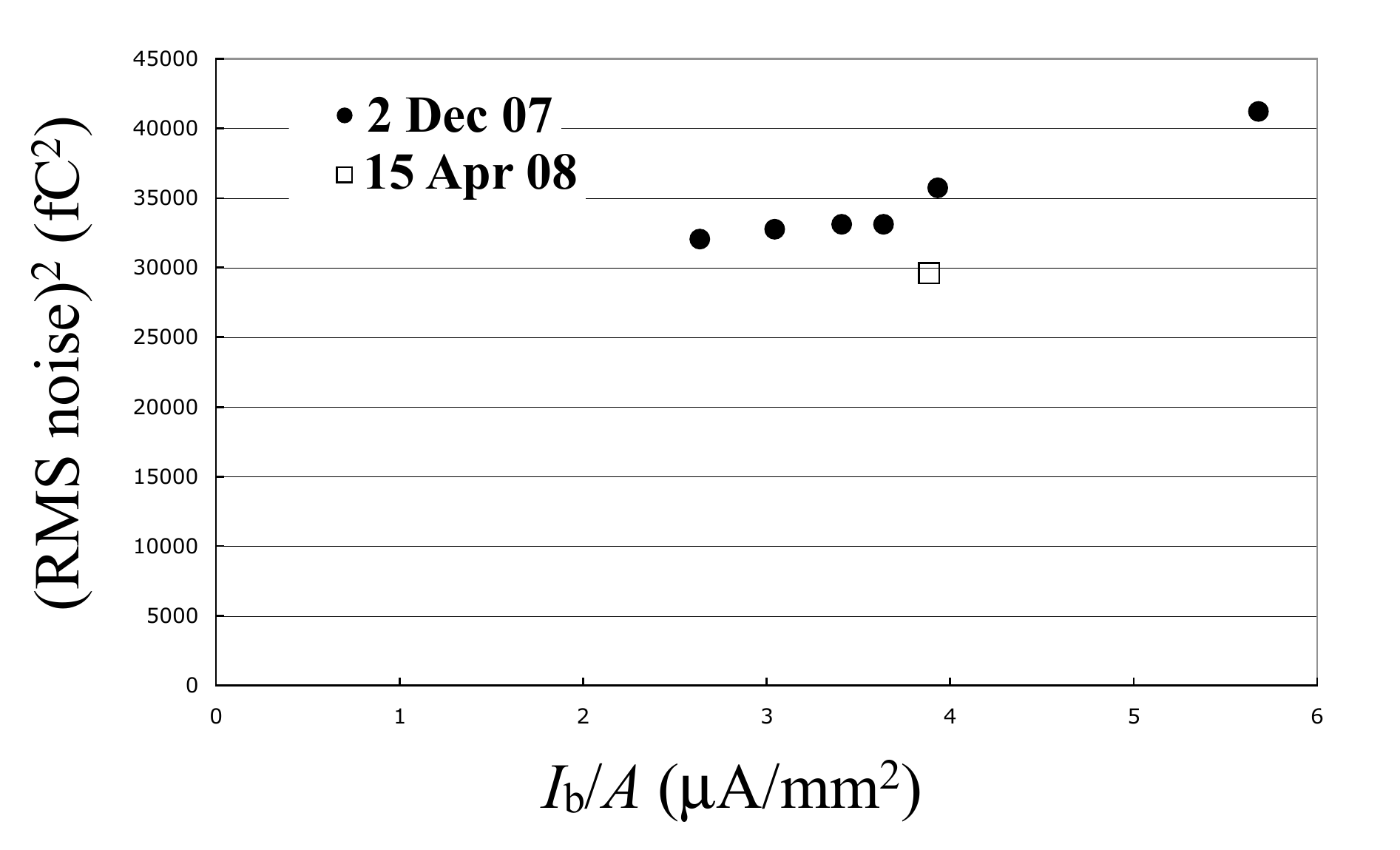}
  \caption{Pedestal rms noise squared {\it vs.} $I_{\rm b}/A$ for CPTA 4.4 mm$^2$ on board 4, for data taken at the time of irradiation, 2 Dec 07 (solid circles) and after room temperature annealing on 15 Apr 08 (open square).}
\label{fig:b7_ch2_rms}
\end{figure}

Figure~\ref{fig:b7_ch2_rms} shows the square of the rms noise as a function of $I_{\rm b}/A$ for CPTA 4.4 mm$^2$ on board 4, for data taken immediately after the irradiation (2 Dec 07).  On 15 Apr 08, 135 days after the irradiation,  the dark current had dropped from 5.7 $\mu$A/mm$^2$ to 3.9 $\mu$A/mm$^2$ for  CPTA 4.4 mm$^2$ on board 4 and from 4.1 $\mu$A/mm$^2$ to 3.0 $\mu$A/mm$^2$ on board 3. The rms of the noise distribution on 15 Apr 08 was about 7\% smaller than at the time of irradiation corresponding to the same leakage current at a fluence of $10^{10}~\rm cm^{-2}$.

\section {FBK 6.2 mm$^2$}

\begin{figure}[htbp]
  \centering
  \includegraphics[width=1\textwidth]{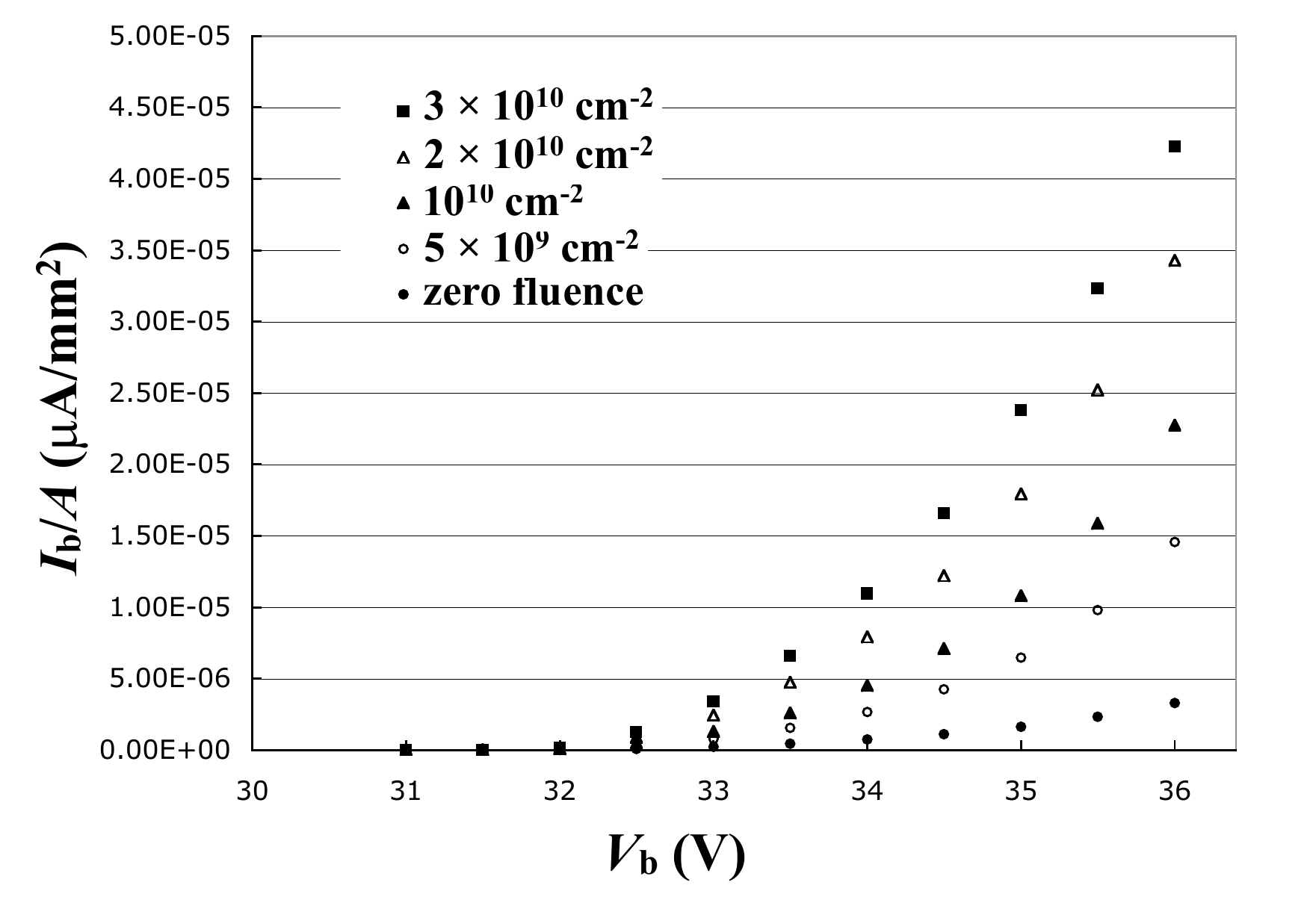}
  \caption{Leakage currents per mm$^2$ for FBK 6.2 mm$^2$ on board 4 as a function of bias voltage for varying proton fluence.}
\label{fig:leak_b7ch3}
\end{figure}

Figure~\ref{fig:leak_b7ch3} shows $I_{\rm b}$ as a function of $V_{\rm b}$ for FBK 6.2 mm$^2$ on board 4 for
fluences of zero, $5\times 10^9$, $10^{10}$, $2\times 10^{10}$, and  $3\times 10^{10}$ protons per cm$^2$. The shape of the $I_{\rm b}$ {\it vs.} $V_{\rm b}$ indicate that the gain {\it vs.} voltage is stable. 
A direct measurement of $MF$ from the mean and width of the response to the 
LED as a function of voltage before and after irradiation shows that the gain 
in the region of nominal voltage varies by about 100 fC/PE per V for FBK 6.2 mm$^2$  on boards 3 and  4.
At a nominal operating voltage of $V_{\rm b}=34~\rm V$, the leakage current increases from 0.8 $\mu$A/mm$^2$ at zero fluence to 10. $\mu$A/mm$^2$ at $3\times 10^{10} \rm ~cm^{-2}$ for board 4.
Similar $I_{\rm b}$ {\it vs.} $V_{\rm b}$ curves were observed for the other  FBK 6.2 mm$^2$ on board 3, where the leakage current at nominal voltage (34 V) increased from  1.2 $\mu$A/mm$^2$ at zero fluence to 5.8 $\mu$A/mm$^2$ at $10^{10} \rm ~cm^{-2}$.

Table 7 shows the values of $I_{\rm b}$, $MF$, and $n_{\rm PE}/F$ and $S/S_0$. The values of $n_{\rm PE}$ are again corrected for the measured deviation of the reference diode (see Table 2). The gain times excess noise factor is observed to be stableat 200 fC/PE for board 1 and decrease from 340 fC/PE to 310 fC/PE for board 4.

\begin{table}[h]
\begin{center}
\caption{Measured properties of the FBK 6.2 mm$^2$ SiPMs. The bias voltage was 34 V. }
\bigskip
\begin{tabular}[]{|c|c|c|c|c|c|}
\hline
{\it Board}   & {\it Fluence \rm ($\rm cm^{-2}$)} & {$I_{\rm b}/A$  ($\mu$A/mm$^2$)} & {$M F$ (fC/PE)}  & { $n_{\rm PE}/F$} & {$S/S_0$} \\
\hline
\hline
{\rm 3} & {zero} & {1.2} & {400} & {180} & {1} \\
\hline
{\rm 3} & {$2.5\times 10^9$} & {2.2} & {400} & {180} & {1.03}\\
\hline
{\rm 3} & {$5\times 10^9$} & {3.3} & {400} & {180} & {1.02} \\
\hline
{\rm 3} & {$7.5\times 10^9$} & {4.2} & {390} & {180} & {0.99} \\
\hline
{\rm 3} & {$10^{10}$} & {5.8} & {400} & {180} & {0.98}  \\
\hline
{\rm 3} & {15Apr08} & {2.8} & {400} & {170} & {0.96} \\
\hline
\hline
{\rm 4} & {zero} & {0.8} & {340} & {170} &{1}\\
\hline
{\rm 4} & {$2.5\times 10^9$} & {1.7} & {330} & {170}& {1.00} \\
\hline
{\rm 4} & {$5\times 10^9$} & {2.6} & {350} & {160}& {0.99} \\
\hline
{\rm 4} & {$7.5\times 10^9$} & {3.6} & {330} & {170}& {0.97} \\
\hline
{\rm 4} & {$10^{10}$} & {4.1} & {320} & {170} & {0.96} \\
\hline
{\rm 4} & {$3\times 10^{10}$} & {10.} & {310} & {150} & {0.84} \\
\hline
{\rm 4} & {15Apr08} & {4.9} & {310} & {170} & {0.93} \\
\hline
\end{tabular}
\end{center}
\label{tab:det}
\end{table}

\begin{figure}[htbp]
  \centering
  \includegraphics[width=1\textwidth]{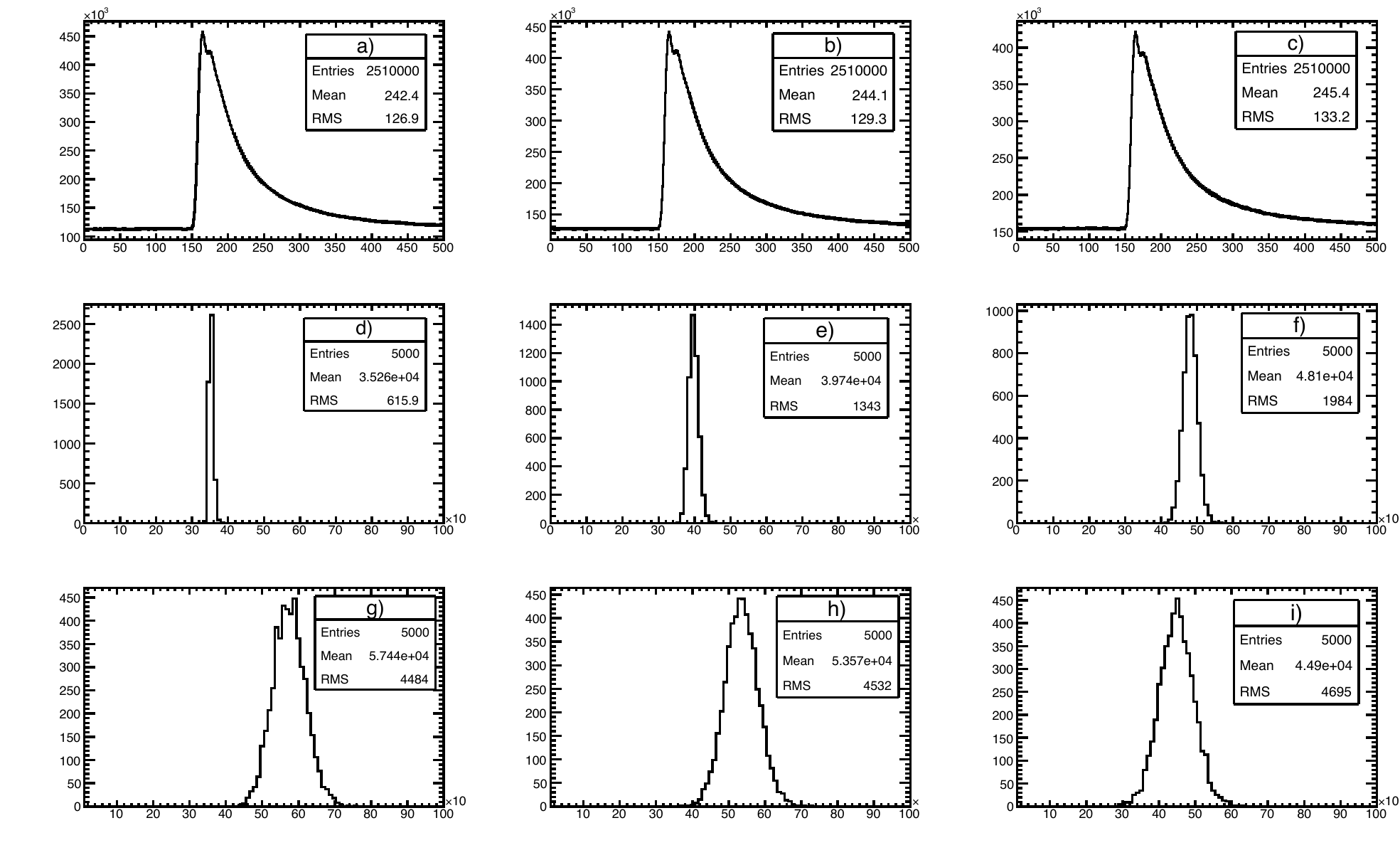}
  \caption{FBK 6.2 mm$^2$  at $V_{\rm b}= 34$ V on board 4:   
  pulse shape a) before irradiation, b) after $10^{10}~\rm cm^{-2}$, and c) after  $3\times 10^{10}~\rm cm^{-2}$; 
  noise distribution d) before irradiation, e) after $10^{10}~\rm cm^{-2}$, and f) after  $3\times 10^{10}~\rm cm^{-2}$; and
   signal distribution in response to LED g) before irradiation, h) after $10^{10}~\rm cm^{-2}$, and i) after  $3\times 10^{10}~\rm cm^{-2}$. 
}
\label{fig:b7ch3}
\end{figure}

The pulse shape for 500 2 ns bins in response to the LED summed over 5000 events is shown in fig.~\ref{fig:b7ch3} for a) zero fluence, b) $10^{10} \rm ~cm^{-2}$, and c) $3\times 10^{10}\rm ~cm^{-2}$ for board 4. The pulse shape was observed to be stable at all fluences on both boards 3 and 4.

The pedestal was summed over 200 ns (bins 1-100 of fig.~\ref{fig:b7ch3} a),b), and c)) to get the noise distributions in fC shown in fig.~\ref{fig:b7ch3} for d) zero fluence, e) $10^{10} \rm ~cm^{-2}$, and f) $3\times 10^{10}\rm ~cm^{-2}$ for board 4. The rms noise increases from 616  fC at zero fluence to 1343 fC at $10^{10} \rm ~cm^{-2}$ to 1984 fC at $3\times 10^{10} \rm ~cm^{-2}$. 

The signal was summed over 200 ns (bins 151-250 of fig.~\ref{fig:b7ch3} a),b), and c)) and the noise was subtracted to get the signal distributions in fC shown in fig.~\ref{fig:b7ch3} for g) zero fluence, h) $10^{10} \rm ~cm^{-2}$, and i) $3\times 10^{10}\rm ~cm^{-2}$ for board 4. The signal on FBK 6.2 mm$^2$ on board 4, relative to  zero fluence and corrected for the reference diode signal, was observed to drop by 4\% at a fluence of $10^{10} \rm ~cm^{-2}$ and 16\% at $3\times 10^{10} \rm ~cm^{-2}$. Similarly, the signal for FBK 6.2 mm$^2$ on board 3 was 2\% lower at  a fluence of $10^{10} \rm ~cm^{-2}$.

\begin{figure}[htbp]
  \centering
  \includegraphics[width=1\textwidth]{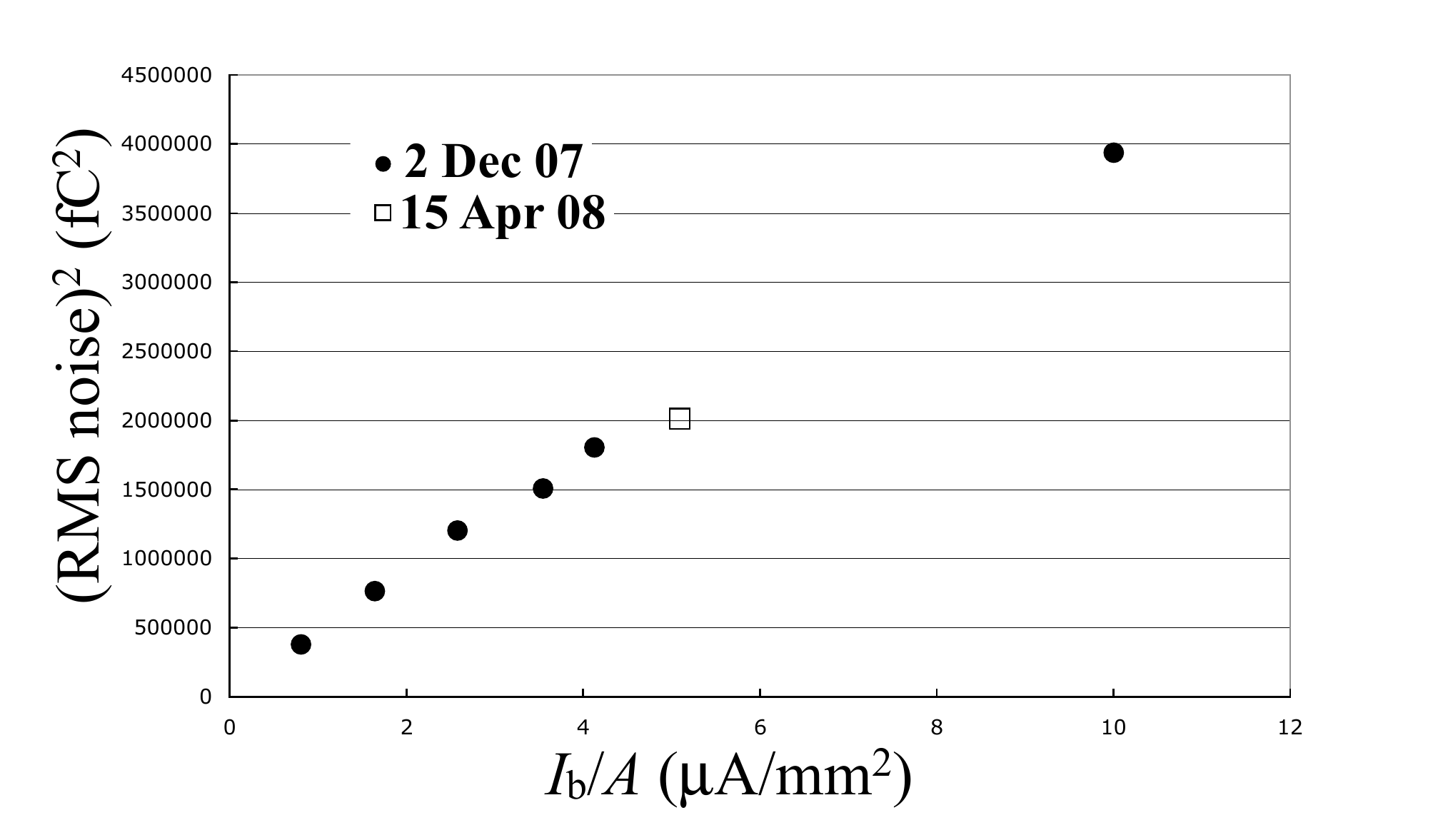}
  \caption{Pedestal rms noise squared {\it vs.} $I_{\rm b}/A$ for FBK 6.2 mm$^2$ on board 4, for data taken at the time of irradiation, 2 Dec 07 (solid circles) and after room temperature annealing on 15 Apr 08 (open square).}
\label{fig:b7_ch3_rms}
\end{figure}

Figure~\ref{fig:b7_ch3_rms} shows the square of the rms noise as a function of $I_{\rm b}/A$ for FBK 6.2 mm$^2$ on board 4, for data taken immediately after the irradiation (2 Dec 07).  On 15 Apr 08, 135 days after the irradiation,  the dark current had dropped from 10. $\mu$A/mm$^2$ to 4.9 $\mu$A/mm$^2$ for  FBK 6.2  mm$^2$ on board 4 and from 5.8 $\mu$A/mm$^2$ to 2.8 $\mu$A/mm$^2$ on board 3. The rms of the noise distribution on 15 Apr 08 was nearly identical to the noise at the time of irradiation corresponding to the same leakage current as interpolated from the measurements at fluences of $10^{10}~\rm cm^{-2}$ and $2\times 10^{10}~\rm cm^{-2}$.

\clearpage

\section {FBK Single Pixel}
\begin{figure}[htbp]
  \centering
  \includegraphics[width=1\textwidth]{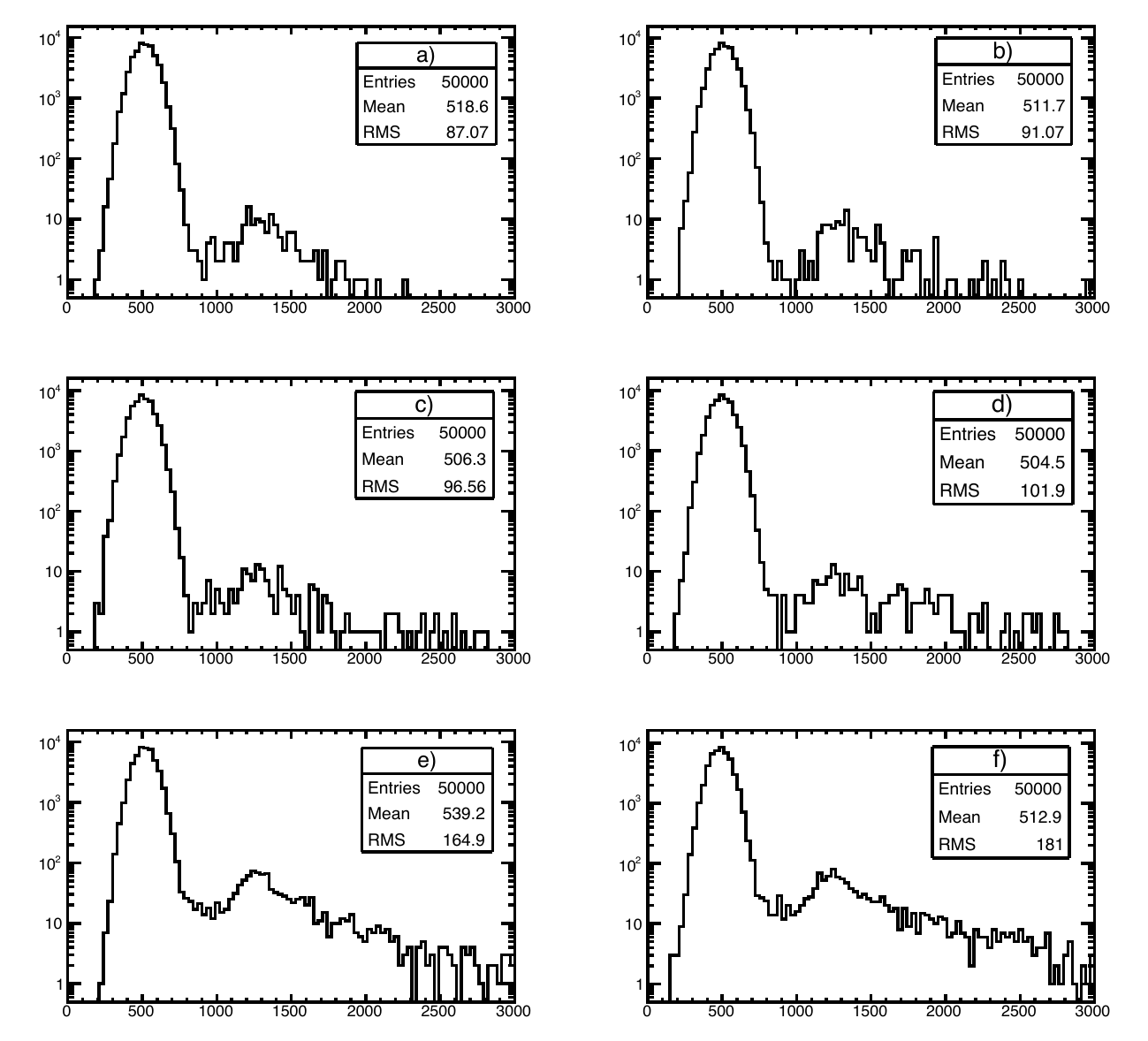}
  \caption{Noise distribution for FBK single pixel  at $V_{\rm b}= 37$ V on board 4:   
  a) zero fluence, b) $2.5\times 10^{9}~\rm cm^{-2}$, c) $5\times 10^{9}~\rm cm^{-2}$,
  d) $10^{10}~\rm cm^{-2}$, e) $2\times 10^{10}~\rm cm^{-2}$ and
  f) $3\times 10^{10}~\rm cm^{-2}$. 
}
\label{fig:single}
\end{figure}

Two of the FBK 6.8 mm$^2$ SiPMs were wired electrically to read out a single 50 $\mu$m pixel. One of these SiPMs (on board 3) developed wire bonding problems prior to the irradiation and is not discussed further. The other single pixel readout (on board 4) was  operated at high gain corresponding to $V_{\rm b}=37 \rm ~V$ to allow detection of single PEs. This SiPM was irradiated to a fluence of $3\times 10^{10}~\rm cm^{-2}$. A total of 10k 1 $\mu$s waveforms were recorded for each partial fluence with no LED, and the pulse height was integrated over 200 ns. The resulting noise distributions are shown in fig.~\ref{fig:single}. Note the data are plotted on a log scale. The location of the single PE peak is seen at approximately 800 fC above the zero PE peak. The leakage current prior to irradiation was 22 nA. At 
 a fluence of $3\times 10^{10}~\rm cm^{-2}$, the leakage current had increased to 150 nA, corresponding to the same order of magnitude value of $I_{\rm b}/A$ as measured in the FBK 1.0 mm$^2$ and 6.2 mm$^2$ SiPM when extrapolated to $V_{\rm b}=37 \rm ~V$. The single pixel becomes slightly noisier with increasing fluence as evidenced by the single PE tail, however, the location of the single PE peak remains stable indicating the gain does not change.

\clearpage

\section{Summary}

\begin{figure}[htbp]
  \centering
  \includegraphics[width=1\textwidth]{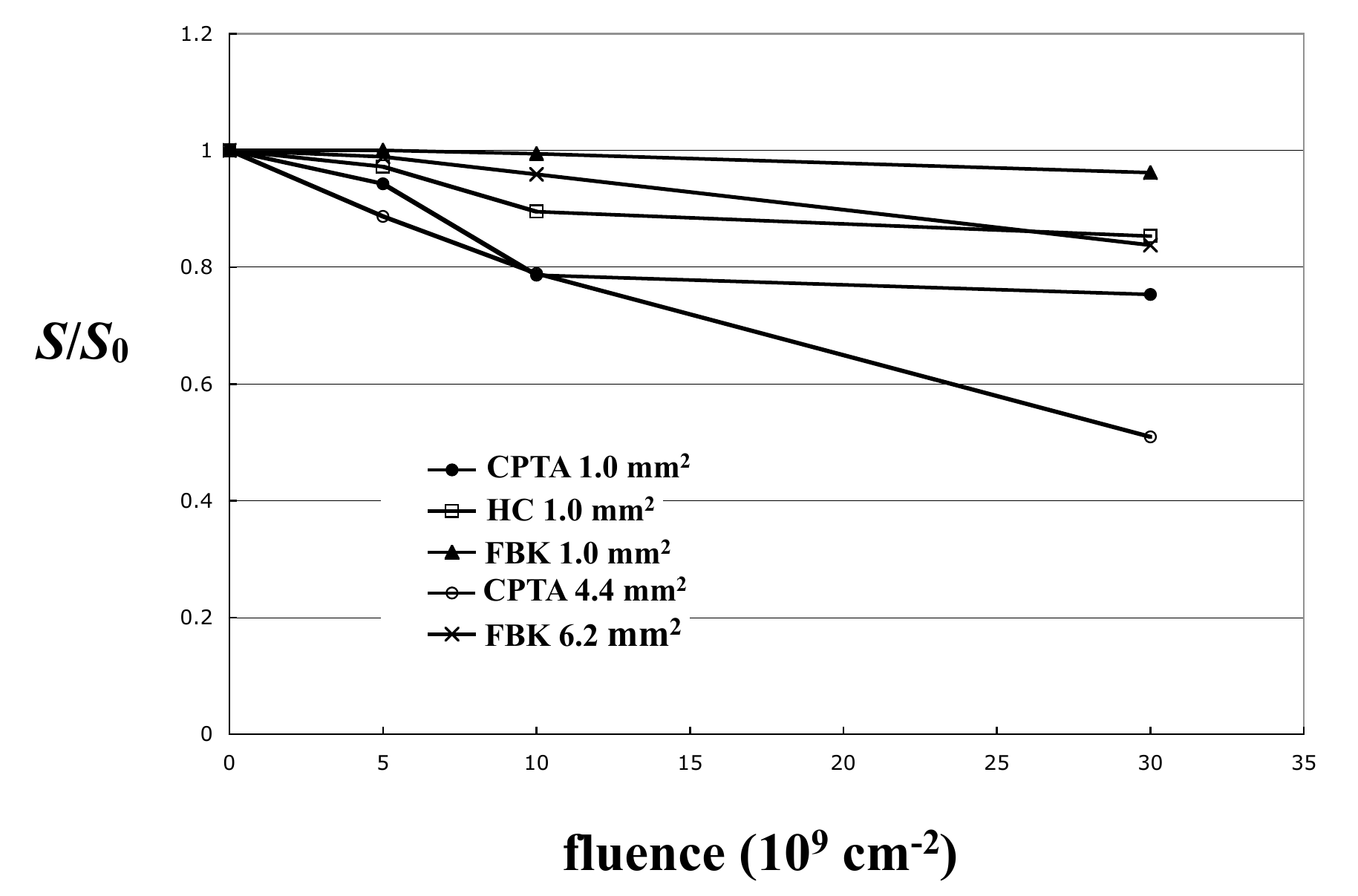}
  \caption{Response of the SiPMs to calibrated LED pulses relative to those at zero fluence ($S/S_0$). The calibration of the LED pulses is done with the reference SiPMs (see Table 2).
  }
\label{fig:signals}
\end{figure}

 We have exposed SiPMs manufactured by 
 Fondazione Bruno Kessler  (1 mm$^2$ and 6.2 mm$^2$), Center of Perspective
Technology and Apparatus  (1 mm$^2$ and 4.4 mm$^2$), and Hamamatsu Corporation  (1 mm$^2$)
using a beam of 212 MeV protons at Massachusetts General Hospital in Boston, MA.
The SiPMs
received fluences of up to $3 \times 10^{10}$ protons per cm$^2$  at operating voltage.
Leakage currents were read continuously during the irradiation, providing a good monitor of the condition of the SiPMs.
The leakage current is found to increase in proportion to the mean square deviation of the noise distribution, indicating the dark counts are due to increased random individual pixel activation. 
At large values of bias currents, the gains are observed to drop due to a lowering of $V_{\rm b}$ due to the voltage drop across the 2 k$\Omega$ input resistor. There is no evidence for any increase in the excess noise factor with irradiation.
Signals in response to calibrated LED pulses (fig.~\ref{fig:signals}) drop by 25\% for CPTA 1.0 mm$^2$, 15\% for HC 1.0 mm$^2$, 4\% for FBK 1.0 mm$^2$, 49\% for CPTA 4.4 mm$^2$, and 16\% for FBK 6.2 mm$^2$ SiPMs after exposure to $3\times 10^{10}$ protons per cm$^{-2}$. For the FBK and HC SiPMs, the reduction in signal is largely attiributed to the reduced gain under large bias currents.
The larger drop for the CPTA SiPMs, especially the 4.4 mm$^2$ CPTA (fig.~\ref{fig:signals}), can be explained by the large dead time caused by the very large quenching resistor, resulting in a $\mu$s deadtime for each pixel. 
 In spite of the drop in signals, all of the SiPMs remained fully functional as photon counters, albeit with increased noise due to increases in dark counts.
The SiPMs are found to anneal at room temperature with a reduction in the leakage current by a factor of 2 in about 100 days.

\section{Acknowledgment}
We acknowledge support of the U.S. National Science Foundation.


\clearpage

\begin{thebibliography}{9}
\bibitem{sipm1}
A. Heering {\it et al.}, ``Performance of Silicon Photomultiplers with the CMS HCAL Front-End Electronics,"
Nucl. Instrum. Meth. A 576 (2007) 341.

\bibitem{sipm2}
A. Heering {\it et al.}, ``Large-Area SiPMs for the CMS Hadron Outer 
Calorimeter," 2007 IEEE Nuclear Science
Symposium Conference Record, Vol. 2, 1545 (2007).

\bibitem{cms}
S. Chatrchyan {\it et al.},
``The CMS experiment at the CERN LHC," JINST 3 (2008) S08004.

\bibitem{lhc}
``LHC Machine,"
Lyndon Evans and Philip Bryant (editors),
JINST 3 (2008) S08001.

\bibitem{lind}
G. Lindstrom, ``Radiation Damage in Silicon Detectors," NIM A512 (2003) 30.

\bibitem{mika}
M. Huhtinen, ``Radiation Environment Simulations for the CMS Detector," 
CERN CMS TN/95-198.

\bibitem{mika2}
M. Huhtinen, ``Optimization of the CMS forward shielding," CMS Note 2000/068. 

\bibitem{slhc} J. Rohlf, "Super-LHC: The Experimental Program,"  talk given at the International Workshop on Future Hadron Colliders (2003)
http://conferences.fnal.gov/hadroncollider/talks/rohlf.pdf

\bibitem{slhc2} J. Freeman, "Calorimeters for the SLHC and VLHC,"  talk given at the International Workshop on Future Hadron Colliders (2003)
http://conferences.fnal.gov/hadroncollider/talks/freeman.pdf

\bibitem{julie}
J. Whitmore, ``System Level Radiation Validation Studies for the CMS HCAL Front-End Electronics ,"
FERMILAB-Conf-03/316-E (2003).

\bibitem{rd50}
D. Bortoletto, ``Recent Results from RD50," 
Nucl. Instrum. Meth. A 569 (2006) 69.

\bibitem{mgh}
E. W. Cascio {\it et al.}, ``The Proton Irradiation Program at the Northeast
Proton Therapy Center," IEEE Radiation Effects Data Workshop, 141 (2003).




\end{thebibliography}
\end{document}